\documentclass[twocolumn,showpacs,preprintnumbers,amsmath,amssymb,superscriptaddress]{revtex4-1}

\usepackage{graphicx}% Include figure files
\usepackage{dcolumn}% Align table columns on decimal point
\usepackage{bm}% bold math
\usepackage{amsthm,amsmath}
\usepackage[x11names]{xcolor}
\usepackage{longtable}
\usepackage{tikz}
\usepackage[T1]{fontenc}
\usepackage{lmodern}
\usepackage{tabularx}
\usepackage{longtable}
\usepackage{array}

\tikzstyle{format}=[draw, thin, rounded corners=5pt, line width =0.5, fill=concepts!10!white]
\usetikzlibrary{arrows,decorations.pathmorphing,fit,matrix,
                patterns, mindmap, trees, positioning, shadings, fadings, 
			    calc, backgrounds}

\colorlet{mindmap1}{SpringGreen4}
\colorlet{mindmap2}{SpringGreen4!85}
\colorlet{mindmap3}{SpringGreen4!70}
\definecolor{elements}{RGB}{53,0,53}
\definecolor{binaries}{RGB}{0,153,0}
\definecolor{n4z}{RGB}{58,218,217}
\definecolor{n3z}{RGB}{13,72,205}
\definecolor{n2z}{RGB}{205,25,13}

\begin{document}

\preprint{}

\title[MSE]{Test set for materials science and engineering with user-friendly graphic tools for error analysis: Systematic benchmark of the numerical and intrinsic errors in state-of-the-art electronic-structure approximations}

\author{Igor Ying Zhang}
\email{igor_zhangying@fudan.edu.cn}
\affiliation{Fritz-Haber-Institut der Max-Planck-Gesellschaft, Faradayweg 4-6, 14195 Berlin, Germany}%
\affiliation{Department of Chemistry, Fudan University, Shanghai 200433, China}%
\affiliation{MOE Key Laboratory of Computational Physical Science, Fudan University, Shanghai 200433, China}%

\author{Andrew J. Logsdail}
\affiliation{Cardiff Catalysis Institute, School of Chemistry, Cardiff University, Cardiff, UK}

\author{Xinguo Ren}
\affiliation{Key Laboratory of Quantum Information, University of Science and Technology of China, Hefei 230026, China}%

\author{Sergey V. Levchenko}
\affiliation{Fritz-Haber-Institut der Max-Planck-Gesellschaft, Faradayweg 4-6, 14195 Berlin, Germany}%

\author{Luca Ghiringhelli}
\email{ghiringhelli@fhi-berlin.mpg.de}
\affiliation{Fritz-Haber-Institut der Max-Planck-Gesellschaft, Faradayweg 4-6, 14195 Berlin, Germany}%

\author{Matthias Scheffler}
\affiliation{Fritz-Haber-Institut der Max-Planck-Gesellschaft, Faradayweg 4-6, 14195 Berlin, Germany}%

\date{\today}

\begin{abstract}
Understanding the applicability and limitations of electronic-structure methods needs careful 
and efficient comparison with accurate reference data. 
Knowledge of the quality and errors of electronic-structure calculations is crucial to advanced
method development, high-throughput computations and data analyses.
In this paper, we present a test set for computational materials science and engineering (MSE), 
that aims to provide accurate and easily accessible crystal properties for a hierarchy of 
exchange-correlation approximations, ranging from the well-established mean-field 
approximations to the state-of-the-art methods of many-body perturbation theory. 
We consider cohesive energy, lattice constant and bulk modulus as 
representatives for the first- and second-row elements and their binaries with cubic crystal 
structures and various bonding characters. A strong effort is made to push the borders of 
numerical accuracy for cohesive properties as calculated using the local-density approximation 
(LDA), several generalized gradient approximations (GGAs), meta-GGAs and hybrids in 
\textit{all-electron} resolution, and the second-order M\o{}ller-Plesset perturbation theory
(MP2) and the random-phase approximation (RPA) with frozen-core approximation based on 
\textit{all-electron} Hartree-Fock, PBE and/or PBE0 references.
This results in over 10,000 calculations, which record a comprehensive convergence test with respect to numerical parameters for a wide range of
electronic structure methods within the numerical atom-centered orbital framework.
As an indispensable part of the MSE test set, a web site is established 
\href{http://mse.fhi-berlin.mpg.de}{\texttt{http://mse.fhi-berlin.mpg.de}}. This
not only allows for easy access to all reference data, but also provides user-friendly graphical 
tools for post-processing error analysis.

\end{abstract}

%\pacs{ }% PACS, the Physics and Astronomy
                             % Classification Scheme.
%Use showkeys class option if keyword
                              %display desired
\maketitle

\section{Introduction}
First-principles electronic-structure calculations have become an indispensable complement
to experiments in physics, chemistry, and materials science, etc. 
Since the exact description of interacting electrons and nuclei is intractable for most systems, 
it is a perpetual challenge to find an ``optimal'' approximation that reconciles 
computational accuracy, efficiency, and transferability across different chemical 
environments and dimensionality. 
In quantum chemistry for atoms and molecules, test sets with accurate reference values for 
various relevant chemical and physical properties
have been since long established~\cite{curtiss:2005A,truhlar:2005A,truhlar:2005B,
truhlar:2005C,truhlar:2005D,truhlar:2014A, hobza:2006A,hobza:2011A, martin:2008A,goddard:2009A,jiang:2012A,xu:2008A,xu:2009A,
grimme:2010A,grimme:2011A}. These test sets play an instrumental role in the development of hierarchical electronic-structure approximations for 
both wave-function theory (WFT) and density-functional theory (DFT). In particular, they are needed 
for validating numerical implementations~\cite{head-gordon:2005A, 
sherrill:2010A,ren:2012A,mahler:2013A}, investigating the basis set convergence~\cite{xu:2010A,truhlar:2012A,head-gordon:2012A, 
furche:2012A,igor:2013A}, and benchmarking the intrinsic limitations of 
the various quantum-chemistry methods~\cite{curtiss:1997A,curtiss:2000A,
curtiss:2005A,xu:2004A,xu:2004B,truhlar:2005A,truhlar:2005B,truhlar:2005C,truhlar:2005D,truhlar:2014A,grimme:2006A,xu:2007A,xu:2009B,xu:2011A,
head-gordon:2009A,ren:2011A,ren:2013A,yang:2013A,furche:2013A,xu:2013A,grimme:2010A,grimme:2011A}. 
For most existing test sets, the developers provide all essential information for each calculation, 
including molecular geometry, basis set, and/or code-specific numerical setting~\cite{gmtkn30,m15db}. 

Condensed-matter physics and materials science is lacking behind, so far, with respect to
comparable benchmark datasets.
Two types of accurate reference data are crucial. One is essentially the exact values, 
obtained from either precise experiments or high-level theoretical calculations. 
They are prerequisite for an unbiased benchmark of the intrinsic errors associated with,  
for example, the exchange-correlation (XC) approximations in DFT or more generally 
the treatment of relativistic effects.
The other type of reference data can help to quantify the numerical error in calculated 
values using any chosen method, 
which might come from the basis set incompleteness, finite $\boldsymbol{k}$-point sampling, 
or other approximations made in the numerical implementation. 
%For a newly proposed method, a comprehensive understanding of the numerical error is 
%important for achieving a correct insight into the intrinsic error
%and showing the right usage of the method as well. 
Unfortunately, for condensed-matter systems, neither of these two types of reference data are 
easy to obtain.
We also mention here the Born-Oppenheimer approximation that decouples the dynamics
of nuclei and electrons. This should be assessed as well, but will not be done in this paper.
%, because of the well-known numerical difficulties in periodic boundary conditions that will be discussed in detail later.
%Still lacking is a general paradigm with well-organized test sets that provide numerically 
%accurate reference data, for a hierarchy of 
%electronic-structure approaches, which would facilitate progress towards the exact 
%solutions from theory.

The paper is structured as follows: After a detailed survey of the test sets that are widely 
used in quantum chemistry, as well as in materials science, we discuss the underlying challenges
to obtain numerically accurate reference data in the latter. 
In Section~\ref{Sec:MSE}, a representative test set for materials science and engineering (MSE) 
is established. It is followed by an overview of a dedicated web site 
\href{http://mse.fhi-berlin.mpg.de}{\texttt{http://mse.fhi-berlin.mpg.de}}.
We take three examples to demonstrate that the web site features a multi-mode access framework, versatile visualization, 
and a linear regression tool for post-processing data analysis.
Section~\ref{Sec:reference} presents the numerical strategies applied in this work to obtain 
the numerically well-converged data on a diverse levels of theory. 
We make our conclusions and present outlooks in Section~\ref{Sec:conclusions}.

\subsection{Test sets in quantum chemistry}
% Representative test sets with accurate reference data are crucial for developing simulation techniques in diverse fields of computational science. 
% In quantum chemistry, a variety of such test sets has been well established~\cite{curtiss:2005A,truhlar:2005A,truhlar:2005B,
% truhlar:2005C,truhlar:2005D,truhlar:2014A, hobza:2006A,hobza:2011A, martin:2008A,goddard:2009A,jiang:2012A,xu:2008A,xu:2009A,
% grimme:2010A,grimme:2011A}, which plays an instrumental role in the development of hierarchical electronic-structure approximations for 
% both wavefunction and density-functional theory. In particular, they are needed for validating the numerical implementation~\cite{head-gordon:2005A, 
% sherrill:2010A,ren:2012A,mahler:2013A}, investigating the basis set convergence~\cite{xu:2010A,truhlar:2012A,head-gordon:2012A, 
% furche:2012A,igor:2013A}, and benchmarking the intrinsic limitations of these quantum chemistry methods~\cite{curtiss:1997A,curtiss:2000A,
% curtiss:2005A,xu:2004A,xu:2004B,truhlar:2005A,truhlar:2005B,truhlar:2005C,truhlar:2005D,truhlar:2014A,grimme:2006A,xu:2007A,xu:2009B,xu:2011A,
% head-gordon:2009A,ren:2011A,ren:2013A,yang:2013A,furche:2013A,xu:2013A,grimme:2010A,grimme:2011A}.

The importance of reliable test sets for the success of quantum-chemistry methods was first 
realized by Pople and 
co-workers~\cite{curtiss:1989A,curtiss:1991A,curtiss:1998A,curtiss:2000A,curtiss:2005A,curtiss:2007A}.
Along with the development of the Gaussian-$n$ theories~\cite{curtiss:1989A,curtiss:1991A,curtiss:1998A,curtiss:2000A,curtiss:2005A,curtiss:2007A},
a hierarchy of extrapolating levels of correlation and basis sets have been developed to obtain 
increasingly accurate thermochemistry. A sequence of pioneering test sets
for quantum chemistry were also developed and used for methdological validation. 
These testsets are now widely recognized as the G$n$ test sets, addressing atomization 
energies and other energetic properties of molecules with increasing numbers of accurate reference
values obtained from experiments~\cite{curtiss:1989A,curtiss:1991A,curtiss:1998A,
curtiss:2000A,curtiss:2005A,curtiss:2007A}. 
One of these test sets, G3/99\cite{curtiss:2000A} generated in 1999, has been widely used 
in the development of density-functional approximations (DFA)
to describe covalent bonding in the main group molecules \cite{curtiss:1997A,curtiss:2000A,curtiss:2005A,xu:2004A,xu:2004B,grimme:2006A,
xu:2007A,xu:2009B,xu:2011A,head-gordon:2009A,ren:2011A,ren:2013A,yang:2013A,furche:2013A,xu:2013A}.
Since then, many well-established test sets have been proposed for different, 
mainly ground state properties of molecules and molecular processes~\cite{curtiss:1991A,curtiss:2000A,curtiss:2005A,truhlar:2005A,truhlar:2005B,truhlar:2005C,truhlar:2005D,truhlar:2014A,
 hobza:2006A,hobza:2011A, martin:2008A,goddard:2009A,jiang:2012A,xu:2008A,xu:2009A,grimme:2010A,grimme:2011A}.
For example, Hobza and co-workers designed the S22 set, comprising 22 non-covalent binding complexes 
of biologic relevance, which can be further divided into 7 systems for hydrogen bonds, 8 for dispersion 
bonds, and 7 for mixed bonds~\cite{hobza:2006A}. Moreover as one of the most popular standard 
benchmark sets, the BH76 set proposed by the Truhlar group consists of forward and reverse barrier 
heights of 19 hydrogen transfer reactions, 6 heavy atom transfer 
reactions, 8 nucleophilic substitution reactions, and 5 unimolecular and association reactions~\cite{truhlar:2005C}. Most recently, the Grimme group
compiled  a comprehensive benchmark test set, GMTKN30, which includes 30 subsets collected from the literature, covering a large section of chemically 
relevant properties of the main-group molecules~\cite{grimme:2010A,grimme:2011A}.

It may appear plausible that reference data, i.e. the accurate values of the relevant chemical and 
physical properties of atoms and molecules, can be acquired from 
experiments~\cite{curtiss:1991A,curtiss:2000A,curtiss:2005A,truhlar:2005A,
truhlar:2005B,truhlar:2014A,jiang:2012A, xu:2008A,xu:2009A}, but the required accuracy 
is not always achievable. The influence of electron-vibrational coupling is often unclear, which,
can hamper the comparison with the directly computed values.
%In the absence of trustworthy experimental reference data, one must resort to references from 
Thus reference data would ideally be created by accurate quantum chemistry methods such as
the coupled-cluster (CC) method with single, double, and perturbative triple excitations, 
CCSD(T)~\cite{sinanoglu:1962A, purvis_full_1982, raghavachari_fifth-order_1989}, or
single, double, triple and perturbative quadruple excitations, CCSDT(Q),
or full configurational interaction (full-CI) theory extrapolated to the complete basis 
set (CBS) limit~\cite{booth:2009A,booth:2010A,booth:2011A}.  For the test sets aforementioned, 
the corresponding reference data are either carefully chosen from accurate 
experiments \cite{curtiss:1991A,curtiss:2000A,curtiss:2005A,truhlar:2005A,truhlar:2005B,
truhlar:2014A,jiang:2012A, xu:2008A,xu:2009A} 
or obtained from high-level first-principles calculations \cite{truhlar:2005C,truhlar:2005D,
hobza:2006A,hobza:2011A,goddard:2009A,martin:2008A}.

% %In general, a valuable test set consists of accurate reference values for well-defined and relevant properties of a representative selection of systems. F
% To obtain the first type of reference data, i.e.\ the exact values of the relevant chemical and physical properties for atoms and molecules, accurate
% quantum chemistry schemes like the coupled-cluster (CC) method with single, double, and perturbative triple excitations, CCSD(T)~\cite{sinanoglu:1962A, 
% purvis_full_1982, raghavachari_fifth-order_1989}, or full configurational interaction (full-CI) theory at the complete basis set limit can be
% applied~\cite{booth:2009A,booth:2010A,booth:2011A}. Furthermore, very accurate results from experiment
% % , as for example the atomization energy reference values that were carefully chosen for the Gn test sets, 
% are available as well. For the test sets aforementioned, the corresponding reference data are either carefully chosen from accurate 
% experiments \cite{curtiss:1991A,curtiss:2000A,curtiss:2005A,truhlar:2005A,truhlar:2005B,truhlar:2014A,jiang:2012A, xu:2008A,xu:2009A} 
% or obtained from high-level first-principles calculations \cite{truhlar:2005C,truhlar:2005D,hobza:2006A,hobza:2011A,goddard:2009A,martin:2008A}.

%An implied prerequisite for the success of any computed test set is that it must be easy to 
%obtain numerically well-converged reference data at different theoretical levels. 
%In quantum chemistry, the basis set incompleteness in real space leads to the main source of 
%the numerical error.
As the most popular choice in quantum chemistry, the atom-centered Gaussian-type orbital (GTO) 
basis sets provide well-converged 
total energies of atoms and molecules with a reasonable basis set size for the mean-field methods,
including the local-density approximation (LDA), generalized gradient approximations (GGAs) and 
hybrid functionals in DFT, and the Hartree-Fock method in WFT. 
%The well-known disability of GTO basis functions to describe the nuclear cusp of the core-electron 
%orbitals~\cite{schwartz:1962A,pack:1966A,morgan:1984A} can be overcome using more compact and flexible 
For an all-electron approach and for heavier atoms (e.g.\ Z>18), i.e.\ when wave-functions 
oscillate more strongly at the core region,
it is more convenient, efficient, and accurate to use numerical atom-centered orbital 
(NAO) basis sets~\cite{delley:1990A,scheffler:2009A}. 
%In consequence, the convergence of total energy in \emph{all-electron} resolution
%can be efficiently approached with the treatment of the core and valence electrons on an equal footing.

%The NAO basis set is more compact and flexible 
For advanced correlation methods which require unoccupied single-particle states, 
e.g.\ the second-order M\o{}ller-Plesset perturbation theory (MP2)~\cite{moller_note_1934}, 
the random-phase approximation (RPA)\cite{bohm_collective_1951, pines_collective_1952, 
bohm_collective_1953,furche:2001A}, or CCSD(T)~\cite{sinanoglu:1962A, purvis_full_1982, 
raghavachari_fifth-order_1989,helgaker:1994A}, the slow basis set convergence is a more
serious problem~\cite{dunning:1989A,klopper:1997A,pulay:2003A,igor:2010A,furche:2012A,igor:2013A}.
It can be ascribed to the inaccurate description of the electron-electron Coulomb 
cusps using smooth orbital product expansions~\cite{klopper:2007A,kong:2012A}.
The so-called correlation-consistent basis sets have been proposed~\cite{dunning:1989A,
igor:2013A}, which allow for an analytic extrapolation to the CBS limit. 
Alternatively by introducing an explicit dependence on the inter-electronic distance into 
the wave function (F12 strategies~\cite{klopper:1991A,klopper:1992A,klopper:2012A,noga:2012A,
kong:2012A}), it is possible to consider the cusp explicitely and suppress the basis set 
incompleteness error at a finite basis set size.
Both techniques were proposed to address the basis-set incompleteness errors in advanced correlation
methods for both atoms and molecules, which have been demonstrated to be
very successful for light elements (e.g.\ Z<18) and for ground-state properties~\cite{dunning:1989A,
igor:2013A,noga:2012A,kong:2012A}. 
Therefore, accurate chemical or physical properties, e.g.\ reaction energies, reaction 
barrier heights and isomerization energies, can be obtained without concern for the 
numerical error cancellation originating from a given finite basis set. 
However, the quantum-chemistry test sets are mainly available for the ground-state properties of 
light elements and small molecules, as it remains a challenge to obtain accurate reference 
data with advanced correlation methods for heavy elements (e.g.\ Z>18), 
excited states, and large systems.

\subsection{Test sets in materials science}
In computational materials science, the situation is more complex and less developed than in the 
quantum chemistry of molecules, and there is an urgent need of accurate test sets for 
the developement of advanced DFAs for solids.
%for there is an increasing demand for accurate and well-established test sets to 
%complement the development and implementation of advanced electronic-structure approaches for solids. 
Staroverov and co-workers considered lattice constants, bulk moduli, and cohesive 
energies of 18 solids, as well as jellium surface energies to benchmark the 
TPSS functional, a meta generalized gradient approximation (meta-GGA)~\cite{tao_climbing_2003}.
They also presented results for LDA, GGA PBE, and meta-GGA PKZB~\cite{staroverov_tests_2004}. 
Heyd and Scuseria used lattice constants and bulk moduli
for 21 solids and band gaps for 8 semiconductors out of this set to investigate the 
screened Hartree-Fock-exchange properties in the HSE hybrid
functional. Comparison was also made with LDA, PBE, and TPSS~\cite{heyd_efficient_2004}.
This test set (or part of it) has been used to benchmark the HSE and HSEsol as implemented
in VASP~\cite{kresse:2006A,kresse:2011A}, to understand the failure of B3LYP for 
solids~\cite{kresse:2007A},
and to develop improved DFAs for solids~\cite{kresse:2009A,gruneis:2010A,kresse:2010A}.
A set of 60 cubic solids was used by Haas, Tran, and 
Blaha to compare the performance of different GGA functionals, LDA and TPSS 
in describing lattice constants and bulk moduli~\cite{haas_calculation_2009,
haas_erratum:_2009,tran_performance_2007}. To compare the accuracy of different van der Waals 
(vdW) functionals applied to solids, Klimes and co-workers determined cohesive properties for
a set of 23 bulk solids~\cite{klimes_van_2011}, based on a set of materials and properties used
by Csonka and co-workers who tested the accuracy of GGA and meta-GGA 
functionals~\cite{csonka_assessing_2009}. Recently, Tran \emph{et al.\ }
performed an extensive test on the lattice constant, bulk modulus, and cohesive energy for 
a large set of XC approximations belonging to rungs 1 to 4 of Jacob's ladder of 
DFT~\cite{tran_rungs_2016}, 
performed in a non-self-consistent way using PBE orbitals. 
The test set used contain 44 strongly and 5 weakly bound solids, most of which were cubic, 
except for hexagonal graphite and h-BN. 
%The calculations were performed in a non-self-consistent way using PBE orbitals for all XCAs.
These test sets of materials with different properties have already been playing an important 
role in developing and benchmarking the DFAs for solids. These test sets exclusively rely on 
reference values from experiment; No quasi-exact calculated results exist, which can 
be used as reference data for solids.
%However, in materials science, temperature and pressure conditions, intrinsic defects and 
%impurities, surfaces, or dislocations can make it difficult to obtain accurate 
%unambiguous reference values from experiment. 

For reasons of comparability and consistency, reference values from theory provide the unique
opportunity to compare calculations based on exactly the same atomic structure and exclude 
finite-temperature, zero-point vibration, relativistic and electron-vibration
coupling effects. 
Recently, the ``gold standard'' method in quantum chemistry, CCSD(T), 
has gained attention in materials science~\cite{gruneis:2013A,booth:2013A,shepherd:2014A,
gruneis:2015A,gruneis:2015B,timothy:2017A}.
A significant step towards an exact description of solids was demonstrated by Booth
\emph{et al.}\ by performing a full-CI quality
calculation with the aid of the Quantum Monte Carlo 
(QMC) stochastic strategy~\cite{booth:2013A}. However, the
numerical difficulty and computational complexity for describing the large extent of solids
has strongly limited the applications of these highly accurate quantum chemistry methods to 
only model systems with very few electrons, basis functions, 
and $\boldsymbol{k}$ points~\cite{booth:2012A,gruneis:2013A,booth:2013A,gruneis:2015B}.
Other QMC stochastic strategies, such as Variational Monte Carlo (VMD)~\cite{bressanini:1999A} 
and Diffusion Monte Carto (DMC)~\cite{needs:2002A} feature a low computational scaling 
with respect to the system size and a weak dependence on basis set. 
QMC has been demonstrated to be as accurate as aforementioned quantum chemistry methods, 
e.g.\ CCSD(T), if certain numerical issues could be considered properly, which includes
the fixed-node error and the form and optimizaiton of the trial function~\cite{grossman:2002A,
mitas:2009A,mitas:2016A}.
%These QMC-relevant numerical problems will not be discussed in this paper.
%thus, it remains infeasible to create a test set with numerically well-converged reference data 
%at CCSD(T) and/or full-CI levels for solids.

Unlike the molecular systems investigated by quantum chemistry, the extended periodic 
materials are often simulated in reciprocal space with periodic boundary conditions (PBC) 
for the sake of fully taking advantage of the periodic symmetry.
However, it also introduces extra numerical difficulties: The first Brillouin zone in reciprocal 
space must be sampled by a finite number of $\boldsymbol{k}$ points,
for which different kinds of $\boldsymbol{k}$-mesh have been proposed to provide an
efficient sampling according to different periodic 
symmetries~\cite{chadi:1973A,marcus:1971,baldereschi:1973A,methfessel:1989A,bloechl:1994A}.
Basis set incompleteness error is another big issue: 
The atom-centered GTO-type basis sets that are dominant in
quantum chemistry become cumbersome for condensed matter systems~\cite{gruneich:1998A,cryscor09:2009A}. In particular, large and especialy diffuse GTO basis sets
might cause severe ill-conditioning problems due to the lack of rigorous orthogonality
for GTO basis functions. The plane-wave basis sets are a popular alternative choice for 
solids as they provide an intrinsically improvable description
of the electronic orbitals with only one parameter: the planewave cutoff energy.
However, the delocalized nature of plane waves makes
it inefficient to describe the localized core electrons surrounding atomic nuclei. 
In practice, the pseudo-potential approximation, which removes the core electrons 
from part of calculations, are often needed together with the plane-wave basis sets
in the projector augmented wave (PAW) framework~\cite{hellmann:1935A,schwerdtfeger:2011A}. 
More recently, it has been demonstrated that the use of compact NAO basis 
sets~\cite{scheffler:2009A,levchenko:2015A} or the linearized augmented plane wave (LAPW) 
strategies~\cite{LAPW:2006A} are able to converge the mean-field approximations in all-electron 
and full-potential resolution with relativistic effects explicitly included as opposed to 
implicitly as is the case for pseudopotential methods.

% Depending on the manner of describing the core electrons,
% relativistic effects can be either included explicitly or implecitly in the case of the pseudopotential methods.
% Due to the basis sets employed, the Kohn-Sham (or Hartree-Fock) orbitals and corresponding densities have to be represented either analytically or numerically.

% Such diversity of numerical implementations in periodic boundary conditions facilitates the applicability of the electronic-structure 
% methods to the systems with different numerical complexity. However, if different algorithms cannot give the same result
% for the same system, 
Such diversity of methods leads to a need for a standard data set that is comparable among different 
numerical implementations.
% it is crucial to have well-converged reference data to benchmark the numerical errors behind these implementations. 
This is indeed the purpose of the $\Delta$-value concept introduced by Cottenier 
et al.~\cite{cottenier:2014A,wien2k:2014A}, which is currently focusing on the 
implementation of the PBE method for solids. The numerically well-converged PBE results from the 
all-electron, full-potential LAPW-based program WIEN2k~\cite{wien2k:1999A,wien2k:2014A} are taken 
as reference. To quantify the numerical errors in the description of equations of state, the
$\Delta$-value is defined as the root-mean-squared (RMS) energy difference between the equations of 
state of two codes, averaged over an exhaustive 
test set of crystalline solids, containing all ground-state elemental crystals.

Despite the enormous success achieved by the (semi)-local density-functional approximations, 
e.g. LDA and GGA PBE, there are several notorious failures of these methods. For instance, there 
exist serious self-interaction errors and significant underestimation of vdW interactions for 
these functionals, which demand more sophisticated approximations. 
The meta-GGAs, in which the orbital kinetic energy density is added to the density functional
evaluation, belong to a higher rung of the Jacob's Ladder in DFT. Representative are the 
nonempirical TPSS~\cite{tao_climbing_2003} and the multiparameter empirical M06-L~\cite{truhlar:2008A}.
A recently proposed nonempirical meta-GGA, SCAN, which satisfies 17 known exact constraints 
appropriate to a semilocal functional~\cite{perdew:2015A}, has been found to show 
a big step ahead in accuracy for manifold chemical and physical properties and has attracted
increasing interest in computational materials science.
The so-called hybrid functionals, e.g.\ HSE06~\cite{krukau_influence_2006}, incorporate 
the information of the occupied orbitals in Hartree-Fock-like "exact exchange". 
The next level of complexity is to derive  DFAs using the information of virtual orbitals. 
Two methods at such level, MP2~\cite{moller_note_1934} and RPA~\cite{bohm_collective_1951, 
pines_collective_1952, bohm_collective_1953, ren:2011A,ren:2012A,ren:2012B,ren:2013A}, 
are state-of-the-art in computational materials science~\cite{pisani:2008A,kresse:2010A,
gruneis:2010A,ren:2011A,pisani:2012A,mauro:2012A,mauro:2013A,booth:2013A,gruneis:2015A}.  
The numerical errors in these methods can either be inherited from the aforementioned 
algorithms to solve the one-electron Kohn-Sham (or Hartree-Fock) equations, or arise 
from extra algorithms, such as the choice of the self-consistent Kohn-Sham orbitals 
for the post-processing evaluations~\cite{igor:2016A}; the resolution-of-identity 
technique to handle the two-electron four-center integrals~\cite{whitten:1973A,
dunlap:2000A,ren:2012A,igor:2015A},
and the localization approximations~\cite{scuseria:2001A,pisani:2008A,oschsenfeld:2010A}
to reduce the computational scaling in these advanced correlation methods. 

In this context, it becomes imperative for computational materials science to have a 
representative test set with numerically well-converged reference values at various 
levels of theory. In spirit of the existing quantum-chemistry test sets and the $\Delta$-value 
concept, we introduce in the following our test set for materials science and engineering, 
coined MSE, which is based on results acquired using density functional methods from LDA, PBE, 
PBEsol, SCAN, TPSS, M06-L, HSE06; and state-of-the-art MP2 and RPA. The numerical convergence 
of these methods is investigated in terms of total energy. Cohesive energies, lattice constants 
and bulk moduli are then reported.
A comprehensive understanding of the numerical errors, particularly in MP2 and RPA, is discussed 
in order to aid the community's pursuit of a numerically
stable implementation of CCSD(T) and full-CI for the exact description of solids.

\section{Test set for materials science and engineering (MSE)}
\label{Sec:MSE}
\begin{figure}
    \includegraphics[width=\columnwidth]{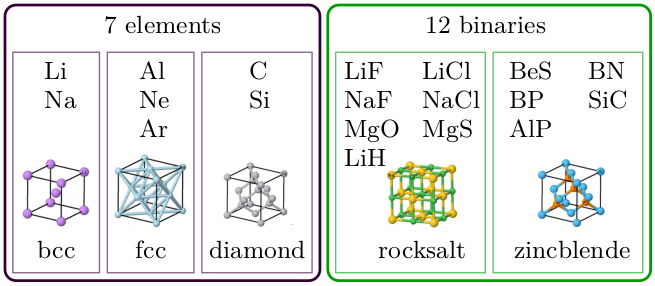}
\caption{The MSE test set containing 7 elements and 12 binaries with cubic structures.}
\label{Fig:MSE}
\end{figure}

%The ground-state crystals of elements in the periodic table are one of natural 
%choices to guarantee the transferability of conclusions drawn from benchmark 
%studies~\cite{cottenier:2014A,cottenier:2016B}. However, whilst elemental 
%crystals provide a diverse range of crystal structures,
%this choice is not representative for binaries, ternaries and others with 
%polar covalent and ionic bonding characters.
In this project, we select 7 elemental solids and 12 binaries with cubic structure,
as the first step in creating the MSE test set. 
As illustrated in Fig.~\ref{Fig:MSE}, the set is composed of elements from 
the first and second rows of the periodic table,
consisting of the body-center-cubic (bcc), face-center cubic (fcc),
diamond, rocksalt, and zincblende structures. Thus, the set 
includes materials with metallic, covalent, ionic, vdW, and mixed 
bonding characters. 

The Fritz Haber Institute ``\textit{ab initio} molecular simulations'' (FHI-aims) electronic-structure package~\cite{scheffler:2009A} is used to generate
the numerically well-converged reference data in the MSE test set. The reasons for this choice are as follows:
\begin{itemize}
	\item The numerical accuracy of the FHI-aims package has been shown to be 
		equivalent to the accuracy of other high-quality codes~\cite{cottenier:2016B}. 
		%the top LAPW codes i.e.\ FHI-aims, WIEN2K and exciting are indistinguishable
		%in accuracy and at the top of all participating codes. 
		In terms of speed, FHI-aims is the clear winner, but this was not considered
		in Ref~\cite{cottenier:2016B}. 
		%Also in a comparision with wavelets, the performance of FHI-aims was excellent.
    \item A range of popular DFAs have been implemented
		in FHI-aims and can be used routinely for atoms, molecules, clusters, and periodic systems. 
		Besides the conventional (semi)-local functionals,  e.g.\ LDA, PBE, PBEsol, SCAN, 
		TPSS and M06-L, a real-space implementation of the exact exchange operator in PBC
		by Levchenko \emph{et al.}\ has been demonstrated to allow for a practical use of 
		hybrid  functionals (including HSE06) and the Hartree-Fock method, as well, for 
		both molecular and periodic systems~\cite{levchenko:2015A}. Furthermore, as will be 
		reported soon, a massively parallel, in-memory implementation of periodic MP2 and RPA 
		methods has also been implemented in FHI-aims using canonical crystalline orbitals.
   \item FHI-aims employs numerically tabulated NAO basis sets~\cite{scheffler:2009A}. The default 
	    basis sets were developed in 2009 starting from a minimal basis that is 
		composed of exact occupied orbitals for spherically symmetric free atoms~\cite{scheffler:2009A}. 
        Such minimal basis captures the essential core-electron behavior in the vicinity of 
		atomic nuclei. Additional ``tiers'' were defined in a step-wise minimization of the LDA 
		total energies of symmetric dimers for all elements. These hierarchical basis sets, 
		\textit{tier-n} ($n$=1--4) for short, can provide the CBS total 
		energies for mean-field approximations (LDA, PBE, PBEsol, SCAN, TPSS, M06-L and HSE06 
		in this project) in \emph{an all-electron description}.
   \item To address the slow basis set convergence of advanced correlation methods (MP2 and RPA 
	    in this project), FHI-aims provides a sequence of NAO basis sets with \emph{Valence 
		Correlation Consistency}~\cite{igor:2013A}, namely NAO-VCC-$n$Z with $n$=2, 3, 4 and 5. 
		The basis set incompleteness error in the valence correlations of MP2 and RPA can be 
		removed using two-point extrapolation schemes~\cite{igor:2013A}. Unless otherwise stated,
		the MP2 and RPA calculations in this work are \emph{valence-only} (frozen-core) using 
		\emph{all-electron} Hartree-Fock and PBE orbitals, respectively. Assuming complete 
		convergence of the $\boldsymbol{k}$-mesh and basis sets, any discrepancy between our 
		results and those obtained with a plane-wave basis should presumably originate from the 
		error of the pseudo-potentials used to generate the valence and virtual orbitals in 
		the self-consistent procedure using Hartree-Fock or PBE. Besides the minimal basis, the 
		NAO-VCC-$n$Z basis sets comprise an additional group of $s,p$ hydrogen-like functions, 
		named enhanced minimal basis~\cite{igor:2013A}. For isolated molecules, such $s,p$ 
		group has been demonstrated to be useful to improve the description of valence 
		correlations~\cite{igor:2013A}. However, the densely packed nature of the condensed-matter 
		systems largely alleviates the difficulty of saturating the $s,p$ basis space for valence 
		correlations. In this work, we exclude the enhanced minimal basis and re-optimize the  
		NAO-VCC-$n$Z basis sets, but do not change the name for simplicity (interested readers 
			are referred to Ref.~\onlinecite{igor:2013A} for the optimization strategy in detail).
\end{itemize}

\begin{figure*}
 \includegraphics[width=2.0\columnwidth]{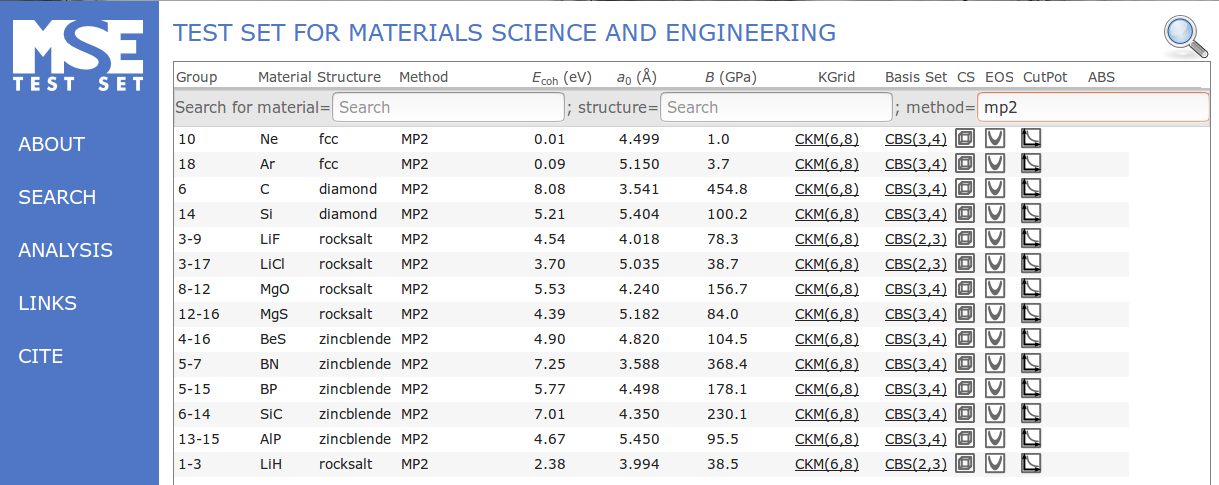}
 \caption{Snapshot of the MP2 data currently available from the MSE web site.}
\label{Fig:mp2}
\end{figure*}

At present the NAO-VCC-$n$Z basis sets in FHI-aims are only available from H to Ar~\cite{igor:2013A}. 
%It remains a challenge to develop
%the correlation consistent basis sets that can accurately converge MP2 and RPA correlations for heavy elements. 
As a consequence, the MSE test set is currently focused on light main group elements (see 
Fig.~\ref{Fig:MSE}). In this paper, we report the numerically well-converged results of MP2 and RPA
for 14 selected materials, including 4 elemental solids (Ne and Ar in the fcc, and C and Si 
in the diamond structure) and 10 binaries (LiF, LiCl, LiH, MgO and MgS with rocksalt, BeS, BN, 
BP, SiC and AlP with zincblende structure).
%MP2 cannot be used to treat metallic systems (here Li, Na, Al) with no energy gap.
%In principle, RPA can handle these systems, but numerical noises are observed in 
%the calculated equations of state, in particular for Al in fcc. This numerical instability
%might be related to the occupation broadening strategy used to find the Fermi level and occupy the Kohn-Sham eigenstates.
%The other three difficult cases for MP2 and RPA are ionic systems (NaF and NaCl), where the frozen-core approximation can not correctly capture the ionic bonding
%characters in these materials. However, to combine the valance-correlation consistent basis sets NAO-VCC-nZ with all-electron RPA and MP2 leads to very wrong 
%cohesive properties. These examples will be discussed in future work. 
We snapshot in Fig.~\ref{Fig:mp2} the MP2 data currently available.
In the long term, the MSE test set shall be extended to include heavy elements and 
non-cubic structures, defects, and surfaces, including representatives
for the majority of systems of interest in materials science and engineering.

Relativistic effects were investigated as well, using the atomic zeroth order regular 
approximation (atomic ZORA)~\cite{scheffler:2009A}. By performing a linear regression 
comparison, we confirm quantitatively that the relativistic effect has a negligible 
influence on the materials formed by light elements, regardless of any XC approximation 
that is used (see the following section for more details).
Thus, we report the reference data without consideration of relativistic effects. 
As a side benefit, this approach excludes the numerical uncertainty arising from different 
relativistic methods~\cite{scheffler:2009A}.
Obviously, for heavy elements, the relativistic effect must be considered carefully.

\section{The web site of the MSE test-set}
\label{Sec:website}
A key feature of our MSE test set project is to establish a dedicated web site 
\href{http://mse.fhi-berlin.mpg.de}{\texttt{http://mse.fhi-berlin.mpg.de}},
which allows for easy access and analysis of the data.
% that record the strong effort made in this work to push the borders of numerical accuracy
% for a hierarchy of electronic structure methods.

\subsection{Multi-mode access to the reference data}
At the MSE home page, an overview of the MSE test set project is given along with a table 
that summarizes all materials available for a given DFA. The materials are sorted 
by their crystalline structures with a layout similar to Fig.~\ref{Fig:MSE}.
The default DFA is PBE; the table for other DFAs is obtained via a drop-down select box. 
% A set of tooptips that contain the reference data of a given material will pop up when you move the mouse over the name of this material in the table.
% When you fly the mouse over a material in the table, a float session will be activated to show the reference data of this material and its convergence benchmark if available.
A search engine allows to access a group of results for a given material, structure, or DFA; 
and/or combination of the above.
Figure~\ref{Fig:mp2} shows a snapshot of the filtered table produced by the search engine, 
listing the MP2 reference data currently available.
This multi-mode search framework, together with a well-organized data structure, guarantees 
easy access to the more than 10,000 calculations in the current MSE test set.

\subsection{Visualization}

The MSE web site provides a quick and easy-to-use visual display of the test set data. 
For any individual reference value calculated for a given material and using a specific DFA, 
the crystal structure and the well-converged equation of state are displayed.
If available, one can also visualize the numerical convergence towards this reference data; 
this includes convergence tests for basis set, $\boldsymbol{k}$-mesh, and internal numerical 
parameters. Furthermore, a statistic analysis of the basis-set convergence for the whole 
test set can be visualised, which is derived using the mean absolute deviation (MAD).

\begin{figure}
    \begin{tikzpicture}[scale=1.0]
    \node[above,align=center] at (0.0cm, 0.0cm) {
	\includegraphics[width=0.8\columnwidth]{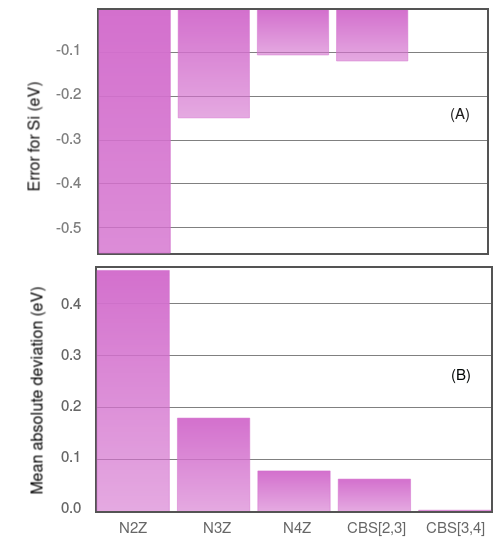}
    };
%     \node[above,align=center] at (0.0cm, -4.0cm) {
% 	\includegraphics[width=0.8\columnwidth]{plots/website/MSE_CE.png}
%     };
 \end{tikzpicture}
 \caption{The basis set incompleteness error in MP2 cohesive energies per atom for (A) Si diamond
	 and (B) 14 materials in terms of mean absolute deviations. 
	 N$n$Z is the short-hand notation of NAO-VCC-$n$Z, and CBS[$n_1$,$n_2$] denotes a complete basis
	 set extrapolation by NAO-VCC-$n_1$Z and $n_2$Z. The CBS[3,4] value is taken as the reference data. 
	 The convergence test is performed with an 4x4x4 $\Gamma$-centered $\boldsymbol{k}$ grid. 
	 In Sec.~\ref{Sec:reference}, more details about the numerical convergence for different methods 
	 will be discussed.
 }
\label{Fig:visualization}
\end{figure}

As an example of this functionality, Fig.~\ref{Fig:visualization} presents the basis set 
incompleteness error in MP2 cohesive energies per atom for (A) Si diamond and (B) 14 
materials representing covalent, ionic, vdW, and mixed bonding characters.
Clearly, NAO-VCC-nZ basis sets provide an improved description for advanced correlation
methods with the increase of the basis set size, which allows for the extrapolation to 
the CBS limit from NAO-VCC-3Z and 4Z (i.e.\ CBS[3,4]).
As a recommendation for practical use, we note that the basis set extrapolation from 
NAO-VCC-2Z and 3Z (CBS[23]) guarantees a near NAO-VCC-4Z quality, but with much less 
computational cost.

% Readers interested in the numerical convergence test with this basis set and method combination or others can easily visualize them online and also download the plots in print quality.
% This visualization function indeed helped us a lot analysis the numerical convergence and locate unexpected numerical spots during the preparation of the accurate reference
% data of nine electronic-structure methods investigated in this MSE test set.

\subsection{Linear regression}

For the benchmark studies of the DFAs, the MAD and the root-mean-square 
deviation (RMSD) are often used to quantify the numerical errors in calculations.
For a given group of materials with the number of materials of $n$ and the targeted observables of $\{Y_i\}$, if the computational data are $\{y_i\}$ with the errors
distributed as $x_i=Y_i-y_i$, we then define the relevant MAD and RMSD as
\begin{equation}
    \begin{split}
         \textrm{MAD}=\frac{1}{n}\sum_{i=1}^{n}\left|x_i\right|\textrm{; RMSD}=\sqrt{\frac{\sum_{i}^{n}x_i^2}{n}}.
         \end{split}
\end{equation}
Despite both MAD and RMSD being measures of accuracy, we note that larger errors have a 
disproportionately large effect on RMSD, making it more sensitive to outliers.
%In the test set project, these outliers might be either numerical anomalies, 
%due to the misoperation when recording the data or even the extreme cases beyond the
%applicability of an electronic-structure method.

A linear regression by means of a least-squares fit allows us to
separate the error into predictable (or systematic) and material-specific 
(or residual) parts~\cite{cottenier:2016A,savin:2015A}. 
The resulting linear model 
\begin{equation}
\hat{L}(y_i)=\beta y_i + \alpha
\end{equation}
allows the analysis of the material-specific deviations $\{\epsilon_i\}$ and the corresponding 
root-mean-square deviation ($\overline{\textrm{RMSD}}$) as
\begin{equation}
  Y_i-\hat{L}(y_i) = \epsilon_i\textrm{; }\overline{\textrm{RMSD}}=\sqrt{\frac{\sum_{i}^{n}\epsilon_i^2}{n}}.
\end{equation}
The systematic error can be determined by the difference between $\textrm{RMSD}$ and 
$\overline{\textrm{RMSD}}$.
It was argued that the material-specific deviations represent a true measure of the 
inadequacy of the method or the numerical incompleteness of the basis set and 
$\boldsymbol{k}$-mesh~\cite{cottenier:2016A,savin:2015A}.
To make the most of this analysis method, we have equipped the MSE web site with a 
linear-regression analysis tool. 
In three examples, we will demonstrate how to gain insight into the numerical and intrinsic 
limitations of a state-of-the-art DFAs.

\emph{(1) Basis set convergence in the MP2 lattice constants}: 
Slow basis set convergence is a well-documented problem for advanced correlation methods 
like MP2 and RPA. Figure~\ref{Fig:mp2} presents the basis set incompleteness errors in MP2 
cohesive energies using the valence correlation consistent basis set in the NAO framework.
Taking the MP2 results in the CBS limit as the reference, 
Table~\ref{Table:linear_regression} lists the RMSDs and $\overline{\textrm{RMSD}}$s of 
the incompleteness errors for NAO-VCC-$n$Z with $n$=2, 3, and 4. 
Despite the basis set errors systematically decreasing with the increase of the basis set size, RMSD and $\overline{\textrm{RMSD}}$ with NAO-VCC-3Z remain about 0.1 \AA, which is
unacceptable for real applications.

\begin{figure}
 \includegraphics[width=1.0\columnwidth]{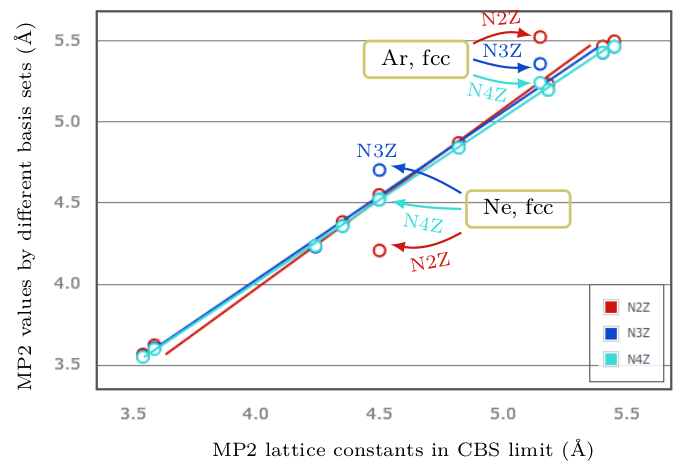}
 \caption{Linear regression of the MP2 lattice constants at different basis sets to the reference 
	 values in the CBS limit. N$n$Z  is the shorthand notation of the valence 
	 correlation consistent basis set NAO-VCC-$n$Z with $n$=2, 3 and 4.}
\label{Fig:linear_regression}
\end{figure}

Figure~\ref{Fig:linear_regression} shows the linear regression analysis. It clearly suggests 
that there are two outliers, which are Ne and Ar with fcc structure. 
We note that Ne and Ar crystals are bonded by vdW forces, which are overestimated
by the MP2 method; however is an intrinsic error of the MP2 method and thus not addressed here.
For isolated molecules, it has been discussed in previous literature~\cite{igor:2013A,ren:2012A}
that an accurate description of weak interactions using advanced DFAs needs either a 
counterpoise correction scheme or very large basis sets with diffuse basis functions to 
address the notorious basis set superposition error.
As suggested by our results in this work, this conclusion is also true for solids governed
largely by vdW interactions. Excluding these two cases from the linear regression analysis 
results in a much better quantitative accuracy (values in parentheses in 
Table~\ref{Table:linear_regression}). 
The RMSD for NAO-VCC-3Z and $\overline{\textrm{RMSD}}$ for NAO-VCC-2Z are below 0.02 \AA, which
clearly suggests that NAO-VCC-3Z or even 2Z with a systematic correction would be good 
enough to converge the MP2 lattice constants for strongly bonded systems,
while the weak interactions must be treated with special care.

\begin{table}
\begin{ruledtabular}
 \caption{RMSDs and material-specific $\overline{\textrm{RMSD}}$ of the basis set incompleteness 
	 errors for the MP2 lattice constants. The values in parentheses exclude Ne and Ar. (unit: \AA)}
 \label{Table:linear_regression}
 \begin{tabular}{ccc}
    Method  & RMSD & $\overline{\textrm{RMSD}}$ \\
    \hline
    N2Z     & 0.148 (0.043) &  0.126 (0.015) \\
    N3Z     & 0.089 (0.017) &  0.071 (0.009) \\
    N4Z     & 0.032 (0.016) &  0.021 (0.007) \\
   \end{tabular}
 \end{ruledtabular}
\end{table}

\emph{(2) Starting point (SP) influence on meta-GGAs and RPA}: 
The common practice of evaluating meta-GGAs, like TPSS, M06-L and SCAN, and advanced many-body 
perturbation methods, like RPA, is by performing an energy evaluation \textit{a posteriori} 
using LDA, GGA or hybrid GGA orbitals~\cite{tran_rungs_2016,ren:2013A}.
Meanwhile, the Hartree-Fock orbitals are used for the MP2 method generally in quantum chemistry.
The self-consistent implementation of meta-GGAs has been realized in many 
computational packages, including FHI-aims.
%However, it remains much more complicated and expensive to perform a self-consistent RPA method, 
%in particular for solids.
%The starting point influence on these methods is of course an unavoidable numerical problem 
%that should be answered precisely. 
In the MSE test set, we provide accurate reference data of TPSS, M06-L and SCAN for both (a) 
self-consistent and (b) non-self-consistent using PBE orbitals, calculations.
We also examine the starting point influence on RPA cohesive energies by using PBE and PBE0 orbitals.
For simplicity and consistency, we use  ``\emph{DFA@SP}'' in this discussion to denote which 
method is based on which starting point (SP) orbitals. 

\begin{table}
\begin{ruledtabular}
 \caption{RMSDs and material-specific $\overline{\textrm{RMSD}}$ of the starting-point 
	 influence on cohesive energies per atom. ``DFA@SP'' denotes which method is 
     based on which starting point orbitals. ``sc-DFA'' denotes the self-consistent
 study. (unit: eV)}
 \label{Table:linear_regression_2}
 \begin{tabular}{ccc}
    Methods to compare  & RMSD & $\overline{\textrm{RMSD}}$ \\
    \hline
    sc-SCAN \emph{vs} SCAN@PBE     & 0.015 &  0.015 \\
    sc-M06-L \emph{vs} M06-L@PBE     & 0.049 &  0.043 \\
    sc-TPSS \emph{vs} TPSS@PBE     & 0.035 &  0.033 \\
    RPA@PBE \emph{vs} RPA@PBE0  & 0.119 & 0.030\\
   \end{tabular}
 \end{ruledtabular}
\end{table}

Table~\ref{Table:linear_regression_2} shows the influence of the starting point on the 
cohesive energies per atom in terms of the RMSD. The linear regression was performed to 
extract the material-specific part of the deviations. Clearly, the starting point influence 
is mild for meta-GGAs: SCAN shows the smallest influence, leading to a RMSD of only 15 meV. 
Meanwhile, the linear regression analysis suggests that these errors are almost 
material-specific. The corresponding systematic errors ($\textrm{RMSD}-\overline{\textrm{RMSD}}$) 
are less than 2 meV. In contrast, RPA is much more sensitive to the choice of the starting point 
orbitals, though the influence is quite systematic:
the material-specific error $\overline{\textrm{RMSD}}$ is only about 30 meV. 
This small error indicates that a careful choice of the starting point orbitals
can improve the RPA performance because of its systematic underestimation 
of cohesive energies~\cite{igor:2016A} and weak interactions~\cite{ren:2011A}.

For meta-GGAs and RPA, the starting point influences on lattice constant and bulk 
modulus are similar: the material-specific deviations are mild, though the 
systematic deviation is quite large for RPA results. The readers interested in this 
topic can easily access the data and perform the linear regression online by themselves. 
It is also easy to investigate the influence of relativistic effects on different
methods and properties in the same manner.

\begin{figure}
 \includegraphics[width=1.0\columnwidth]{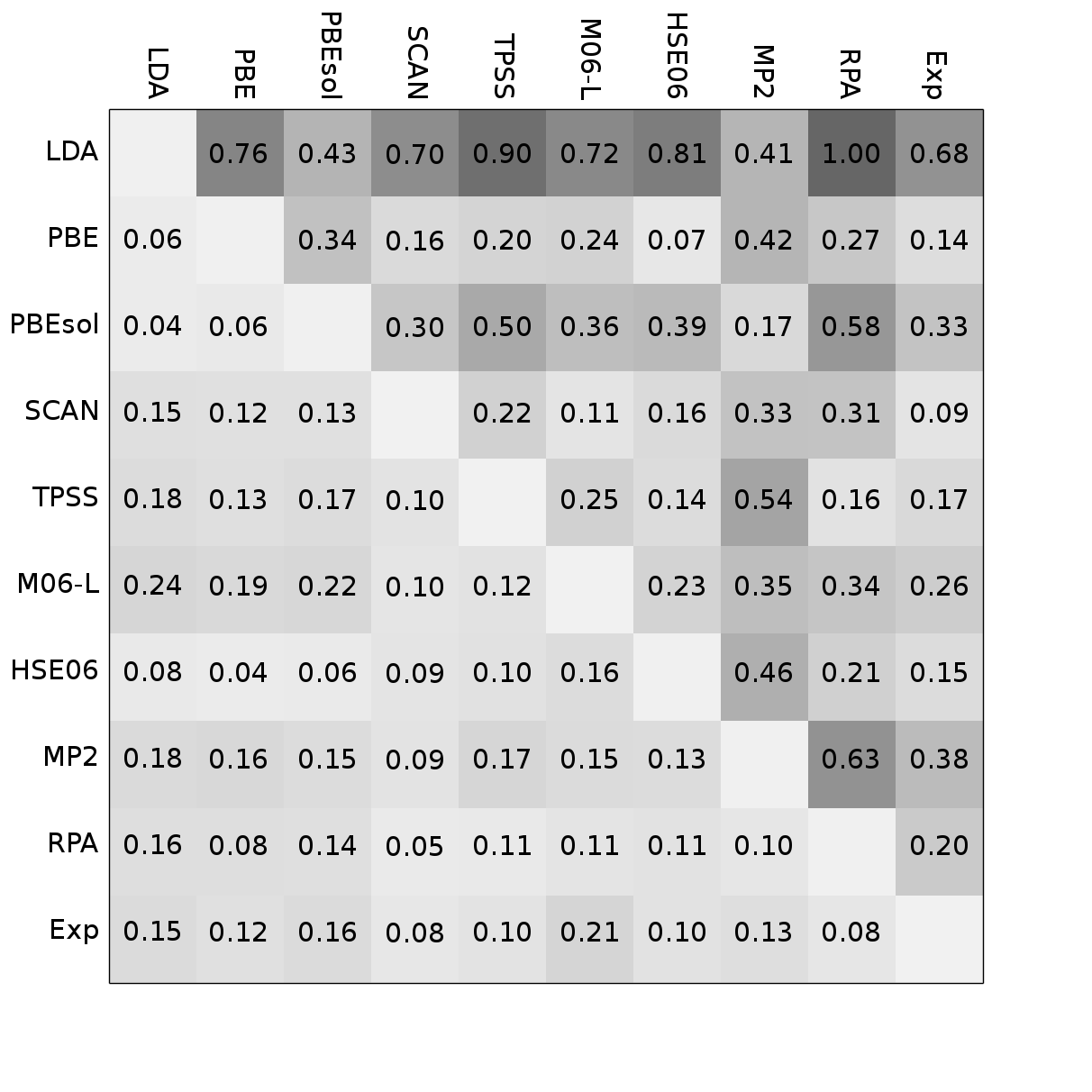}
 \caption{Cross-over comparison of cohesive energies between different methods and 
	 experiment (Exp) in terms of the RMSD. The direct RMSDs are shown in the up 
	 triangular part, while the values in the lower triangular part are the 
     material-specific errors ($\overline{\textrm{RMSD}}$s) after the linear 
     regression. (unit: eV)}
\label{Fig:cross_comparison}
\end{figure}

\emph{(3) Cross-over comparison between different methods}: 
%Benchmark studies of the intrinsic limitations in a given method usually analyse the deviation
%from the exact experimental or theoretical data statistically.
%%, but none of these experimental references can be obtained easily for condensed matter systems. 
%In contrast to the experimental data, reference values from theory distinguish themselves by allowing for a direct comparison of the calculations
%that use the same atomic structure and exclude finite-temperature, zero-point vibration, relativistic and electronic-vibrational coupling effects.
%To provide numerically well-converged reference data at CCSD(T) or full-CI levels is a long-term plan for our MSE test set. 
%In our current scenario, we demonstrate that, in spite of the absence of exact solutions from theory, we can gain insights into the intrinsic limitations 
%of the state-of-the-art methods by a careful cross-over comparison among a hierarchy of electronic-structure methods.
Figure~\ref{Fig:cross_comparison} summarises the cross-over comparison of cohesive 
energies between different methods. RMSDs and the material-specific 
$\overline{\textrm{RMSD}}$s are shown in different subtables. 
In this comparison, RPA@PBE reference data were used and meta-GGAs were calculated self-consistently.
The nine DFAs investigated covers all five rungs of the Jacob's ladder of DFT.
From the many-body perturbation theory point of view, MP2 and RPA consider the electronic 
correlations explicitly in the many-body interaction picture,
while others are all mean-field approximations. 

LDA and RPA are the density functional approximations on the lowest and highest rungs 
of the Jacob's ladder, respectively, and they 
show the largest deviation in the cohesive energy calculations, leading to a RMSD of 
over 1 eV. In fact, LDA shows a large disagreement with all other methods,
with the average RMSDs about 0.7 eV. This observation agrees with a widely accepted 
argument that the local density approximation derived from homogeneous 
electron gas completely misses the high order density derivative or response information 
that is important for a proper description of chemical bonding.
However, our linear regression analysis suggests that this error is quite systematic: 
The material-specific $\overline{\textrm{RMSD}}$ is only 0.16 eV between LDA and RPA.

A similar analysis can be applied to study the intrinsic limitations in MP2 and RPA.
It is well-known that the RPA method performs unsatisfactorily in the calculations of 
cohesive energy for solids~\cite{adrienn:2010A,gorling:2010A,furche:2013A,ren:2013A}. 
From the many-body perturbation point of view, this error is due to the lack of an 
infinite summation of the second-order exchange diagrams, 
which is necessary to eliminate the notorious one-electron self-interaction error in 
the RPA method~\cite{furche:2013A,ren:2013A}.
On the other hand, the MP2 method itself is exactly free from such self-interaction error,
but completely ignores high-order perturbative diagrams.
In consequence, a large RMSD of about 0.68 eV between MP2 and RPA cohesive energies is
observed, which can be traced back to the dissimilarity of underlying 
physical models used in the two methods. Recently, much effort has been devoted to
improve RPA and MP2 from different perspectives, but these approaches often exacerbate
the computational complexity dramatically. Our linear regression analysis suggests
that the difference between MP2 and RPA cohesive energies is also quite systematic,
leading to a material-specific deviation $\overline{\textrm{RMSD}}$ of only about 100 meV.

%Despite the nine electronic-structure methods investigated in this work being proposed
%from different perspectives and for different targets, 
%the cross-over comparison of $\overline{\textrm{RMSD}}$s among them clearly suggests that
%most of the errors can be quite systematic despite their dissimilar origins. 
%As a consequence, it is possible to effectively improve the density functionals by 
%hybridizing these methods in different levels of theory,
%evident by a great success of a new family of functionals, namely the double-hybrid approximations in quantum chemistry~\cite{igor:2011A,igor:2012A}. 
%Based on the material-specific $\overline{\textrm{RMSD}}$s shown in Fig.~\ref{Fig:cross_comparison},
%we argue that this hybrid strategy should work well for solids, but the intrinsic errors in these double hybrid approximations may be approximately 100 meV;
%Any further improvement in accuracy would require a correct understanding of the connection between the material-specific errors and the essential inadequacy in the methods,
%i.e.\ the self-interaction errors and/or the lack of higher order diagrams in correlation.

We also compared the results of different methods with experimental data that were corrected for thermal and zero-point vibrational 
effects~\cite{tran_rungs_2016,cottenier:2014A}. In-line with the observation from the above cross-over comparision, most of the errors in 
different theretical approximations are quite systematic, leading to the material-specific errors $\overline{\textrm{RMSD}}$s in 9 methods
are lower than 210 meV, with M06-L presenting the largest difference. The visualization of the linear regression analysis suggests that abnormally
large errors in M06-L occur in metallic systems (Na and Al), which can be ascribed to an incorrect oscillation of the exchange enhancement factor of M06-L
when approaching to the uniform electron gas limit~\cite{perdew:2013A}. Taking these two points out, 
the RMSD and $\overline{\textrm{RMSD}}$ of M06-L are reduced to 131 meV and 95 meV, respecitvely.

To summarise this section, we took three examples to demonstrate the usage of our MSE web site. 
In the context of ongoing innovation in computational materials science, 
driven by data technology, there is a growing awareness of the importance of effective data 
sharing and recycling. Here, we argue that a 
dedicated test-set web site should be about more than an easy, static access of the reference data. 
In order to liberate the power of test sets, of key importance is providing friendly analysis 
interface that facilitate the users to play with the online data, 
repeat the observation in the original papers, and even gain new insights from the data by themselves.

\section{Numerically well-converged reference data}
\label{Sec:reference}
%We would like to emphasize that  
In this section, we introduce the numerical strategies applied to obtain the reference data in 
the MSE test set. Numerically well-converged reference data at various levels of 
theory are crucial to achieve an unbiased benchmark and discussion in the previous section.

Due to the use of NAO basis sets, the electronic-structure problem is addressed using numerical 
integration in FHI-aims. The specific technical aspects behind the numerical setting of the 
grids used for three-dimensional integrations are described in Ref.~\onlinecite{scheffler:2009A}
for mean-field approximations, and in Ref.~\onlinecite{igor:2013A} for advanced correlation methods.
In this work, very dense grids are employed to ensure an meV accuracy in total energy 
consistently for all methods. 

For hybrid functionals (HSE06 in this work) and more sophisticated treatments of the 
exchange and correlation (e.g.\ MP2 and RPA), the evaluation of electron repulsion 
integrals (ERIs) is a major computational bottleneck. Within FHI-aims, an efficient 
localized resolution-of-identity (called ``RI-LVL'') method~\cite{igor:2015A} was developed,
which expands the basis-function products in a set of auxiliary basis functions located at 
the same atoms as the basis functions. Of key importance is the completeness of the 
auxiliary basis set, which ultimately determines the accuracy of the RI-LVL method.
For isolated molecules, Ihrig \emph{et al.} have demonstrated that the RI-LVL errors in HSE06, 
MP2 and RPA total energies per atom can be converged to sub-meV using a relatively small, 
but sufficient auxiliary basis set~\cite{igor:2015A}.

In FHI-aims, the RI-LVL method has also been adopted in the periodic implementation of 
HSE06~\cite{levchenko:2015A}, MP2 and RPA methods. For all 19 materials in the MSE test set,
we have carefully converged the RI-LVL errors in HSE06, 
MP2 and RPA total energies with respect to the size of the auxiliary basis set. Our results 
demonstrate that the accuracy of the RI-LVL method developed initially for
isolated molecules is completely transferable to extended systems. Since it is a specific
numerical issue for FHI-aims, the results are included in the MSE web site (see 
Sec.~\ref{Sec:website} for usage) and we refer the readers to Ref.~[\onlinecite{levchenko:2015A}] 
and [\onlinecite{igor:2015A}] for more details and discussions about the RI-LVL implementations.

In contrast, the numerical error in terms of $\boldsymbol{k}$-mesh and basis set sizes is a universal difficulty for any software implementation.
In the following section, we focus on the numerical convergence in both aspects for
the total energy per atom of each  method and each material in the MSE test set.
% Besides these converged total energies per atom,
% the cohesive energy, lattice constant and bulk modulus are also reported as the reference data for the MSE test set.

\subsection{The \textbf{\textit{k}}-mesh convergence}

\begin{table*}
\begin{ruledtabular}
 \caption{The $\boldsymbol{k}$-meshes that are used for different methods. 
 ``GC'' denotes an evenly spaced $\Gamma$-center $\boldsymbol{k}$-mesh and ``MP'' for the Monkhorst-Pack one.
 For mean-field approximations, the convergence criteria is 1 meV (or tighter) in the calculations of the total energy per atom. 
 For MP2 and RPA, the complete $\boldsymbol{k}$-mesh extrapolation from $6\times6\times6$ and $8\times8\times8$, denoted as CKM(6,8) is adopted
 with $\alpha=3$ (see Eq.~\ref{Eq:k-mesh-relation}).}%, which converges the total energies per atom of five selected materials within 2 meV.
 \label{Table:k-mesh}
 \begin{tabular}{ccccccccc}
      & & &\multicolumn{2}{c}{(semi-)local DFAs}&\multicolumn{2}{c}{hybrid DFAs}&\multicolumn{2}{c}{MP2 and RPA}\\
	  & Structure& Index & Type & \textit{\textbf{k}}-points & Type & \textit{\textbf{k}}-points & Type & \textit{\textbf{k}}-points\\
  \hline
  Li   &  bcc    	& 01 & GC & $16\times16\times16$ &GC & $16\times16\times16$ & -- & --\\
  Na   &  bcc 	 	& 02 & GC & $16\times16\times16$ &GC & $16\times16\times16$ & -- & --\\
  Al   &  fcc   	& 03 & GC & $20\times20\times20$ &GC & $20\times20\times20$ & -- & -- \\
  Ne   &  fcc 		& 04 & MP & $4\times4\times4$    &GC & $6\times6\times6$    & GC & CKM(6,8)\\
  Ar   &  fcc 		& 05 & MP & $4\times4\times4$    &GC & $6\times6\times6$    & GC & CKM(6,8)\\
  C    &  diamond 	& 06 & MP & $6\times6\times6$    &GC & $10\times10\times10$ & GC & CKM(6,8)\\
  Si   &  diamond 	& 07 & MP & $6\times6\times6$    &GC & $10\times10\times10$ & GC & CKM(6,8)\\
  LiF  &  rocksalt 	& 08 & MP & $6\times6\times6$    &GC & $6\times6\times6$    & GC & CKM(6,8)\\
  LiCl &  rocksalt 	& 09 & MP & $6\times6\times6$    &GC & $6\times6\times6$    & GC & CKM(6,8)\\
  NaF  &  rocksalt 	& 10 & MP & $6\times6\times6$    &GC & $6\times6\times6$    & -- & --\\
  NaCl &  rocksalt 	& 11 & MP & $6\times6\times6$    &GC & $6\times6\times6$    & -- & --\\
  MgO  &  rocksalt 	& 12 & MP & $6\times6\times6$    &GC & $6\times6\times6$    & GC & CKM(6,8)\\
  MgS  &  rocksalt 	& 13 & MP & $6\times6\times6$    &GC & $8\times8\times8$    & GC & CKM(6,8)\\
  BeS  &  zincblende 	& 14 & MP & $6\times6\times6$    &GC & $8\times8\times8$    & GC & CKM(6,8)\\
  BN   &  zincblende 	& 15 & MP & $6\times6\times6$    &GC & $8\times8\times8$    & GC & CKM(6,8)\\
  BP   &  zincblende 	& 16 & MP & $6\times6\times6$    &GC & $10\times10\times10$ & GC & CKM(6,8)\\
  SiC  &  zincblende 	& 17 & MP & $6\times6\times6$    &GC & $10\times10\times10$ & GC & CKM(6,8)\\
  AlP  &  zincblende 	& 18 & MP & $6\times6\times6$    &GC & $8\times8\times8$    & GC & CKM(6,8)\\
  LiH  &  rocksalt 	& 19 & MP & $6\times6\times6$    &GC & $8\times8\times8$    & GC & CKM(6,8)\\
 \end{tabular}
 \end{ruledtabular}
\end{table*}

The \textbf{\textit{k}}-mesh convergence for LDA and PBE methods is investigated together with $\Gamma$-center (GC) and Monkhorst-Pack (MP) sampling strategies~\cite{monkhorst:1976A,
FHI-aims:2015A,VASP:2015A}, both of which span the first Brillouin zone in an evenly spaced manner. The MP \textbf{\textit{k}}-mesh coincides 
with the GC one for an uneven number of grid points~\cite{VASP:2015A,FHI-aims:2015A}. In this project, we choose a set of \textbf{\textit{k}}-meshes
with even numbers of grid points. Our observation is consistent
with previous investigations where the MP \textbf{\textit{k}}-mesh shows faster convergence behavior than the GC \textbf{\textit{k}}-mesh
for insulators and semiconductors at the semilocal density functional level of theory~\cite{VASP:2015A,FHI-aims:2015A}. 
In order to achieve 1 meV accuracy, the $4\times4\times4$ MP grid is sufficient for rare-gas crystals (Ne and Ar with fcc structure), and the $6\times6\times6$ MP
grid for other ionic- and covalent-bond systems. To achieve the same accuracy, an $8\times8\times8$ (sometimes even denser) grid is often necessary with 
the GC $\boldsymbol{k}$-mesh. However, the slow \textbf{\textit{k}}-mesh convergence in the calculation of metallic systems is more serious, and cannot be 
improved using the MP \textbf{\textit{k}}-mesh. 
%{\bf The interested reader can reach the detailed results by vising either our MSE web page or the NoMaD repository (see Sec.~\ref{Sec:data} for usage). Need reframe!}
Based on LDA, PBE, and three meta-GGAs convergence benchmarks, we summarize the \textit{\textbf{k}}-mesh setting in Table \ref{Table:k-mesh}, 
recommended for all semi-local DFAs investigated in this paper: LDA, PBE, PBEsol, TPSS, M06-L and SCAN.

A linear-scaling HSE06 method for condensed matter systems has been implemented in FHI-aims~\cite{levchenko:2015A}. In order to fully utilize the sparsity of 
the density matrix in real space, this HSE06 implementation needs a projection of density matrix in a $\Gamma$-center $\boldsymbol{k}$-mesh to the 
corresponding Born-von-Karman (BvK) supercell by means of Fourier transformation. Furthermore, FHI-aims features a massively parallel, in-memory implementation 
of canonical MP2 and RPA methods for periodic systems. At present, HSE06, MP2 and RPA calculations can be performed only with the $\Gamma$-center
\textit{\textbf{k}}-mesh in FHI-aims. MP2 and RPA calculations have been reported using  $\boldsymbol{k}$-meshes centered at $\Gamma$-point for the 
projector-augmented-wave method and plane wave basis sets~\cite{kresse:2010A,gruneis:2010A,gruneis:2011A,booth:2013A,gruneis:2015A,gruneis:2015B}.

\begin{figure}
 \includegraphics[width=1.0\columnwidth]{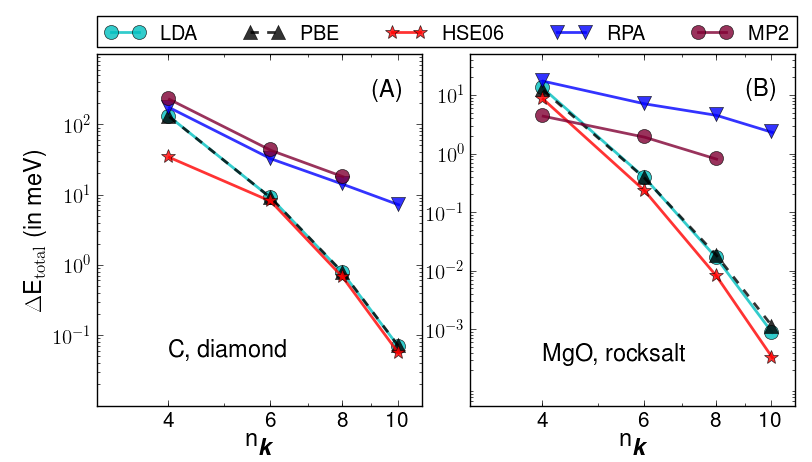}
 \caption{Dependence of the total energy per atom on the number of \textbf{\textit{k}} points $n_{\textbf{\textbf{k}}}$ (along one direction).
 The $\Gamma$-center \textbf{\textit{k}}-mesh is used. C diamond (A) and MgO rocksalt (B) results are presented on a logarithmic scale. 
 The basis sets used are \emph{tier}-2 for LDA and PBE and NAO-VCC-2Z for RPA and MP2. The reference total energies are
 calculated with the $12\times12\times12$ \textbf{\textit{k}}-mesh for LDA, PBE and HSE06. For RPA and MP2, the references are extrapolated by
 CKM(8,10) and CKM(6,8) with $\alpha=3.0$, respectively.}
\label{Fig:k_grid}
\end{figure}

Table \ref{Table:k-mesh} lists the \textit{\textbf{k}}-meshes for converged HSE06 calculations:
Generally speaking, HSE06 shows a similar convergence behavior as LDA and PBE using the $\Gamma$-center \textit{\textbf{k}}-mesh.
Notable differences happen only with very rough grids, e.g.\ in the rare-gas crystals (Ne and Ar) with $2\times2\times2$ and ionic crystals with
$4\times4\times4$. Such abnormality can be ascribed to an improper description of the integrable singularity of the Coulomb potential in
reciprocal space~\cite{kresse:2006A,alavi:2008A}, which is mathematically equivalent to a slow decay of Coulomb interaction in condensed matter systems. 
In FHI-aims, a cut-Coulomb operator is used for HSE06 to generate the Coulomb matrices at every $\boldsymbol{k}$-point~\cite{krukau_influence_2006,levchenko:2015A}. 
However, a reasonably dense $\boldsymbol{k}$-mesh is still required. 
Taking C diamond and MgO rocksalt as examples, Fig.~\ref{Fig:k_grid} shows the \textbf{\textit{k}}-mesh incompleteness error of
total energies per atom for five methods:
\begin{equation}
 \Delta E_{\textrm{total}}[n_{\textit{\textbf{k}}}]=E_{\textrm{total}}[n_{\textit{\textbf{k}}}]-E_{\textrm{total}}[\infty],
\end{equation}
where $n_{\textit{\textbf{k}}}$ is the number of \textbf{\textit{k}} points in each direction ($n_{\textbf{\textbf{k}}}\times
n_{\textbf{\textbf{k}}}\times n_{\textbf{\textbf{k}}}$). Our results reveal that a $\Gamma$-center mesh with $6\times6\times6$ grid points 
is enough for HSE06 to address this singularity issue, thus providing a similar \textit{\textbf{k}}-mesh sampling quality 
as LDA and PBE. %(see supporting information for details).

In the framework of the mean-field approximations, the evaluation of the exchange-correlation energy 
needs only the electron density and/or occupied orbitals. However, the advanced correlation methods, i.e.\ MP2 and RPA, explicitly consider the excitations
that are between occupied and virtual orbitals and that cross over different \textit{\textbf{k}} points.
Therefore, it is not surprising to see that MP2 and RPA calculations for solids have a much higher demand on the
sampling of the first Brillouin zone. The \textit{\textbf{k}}-mesh convergence of the periodic MP2 method has been investigated
with the $\Gamma$-center sampling strategy by Gr\"uneis \emph{et al.}~\cite{gruneis:2009A,gruneis:2010A}.
In FHI-aims, we adopt a hybrid strategy to generate the Coulomb matrices in MP2 and RPA calculations, i.e.\ using the cut-Coulomb operator 
only for the $\Gamma$ point and the full Coulomb operator for the rest of the $\boldsymbol{k}$ points. For MP2 and RPA, this hybrid choice shows a faster $\boldsymbol{k}$-mesh convergence compared to 
the cut-Coulomb strategy used in HSE06 calculations~\cite{levchenko:2015A}.
As Fig.~\ref{Fig:k_grid} illustrates, the total energies per atom calculated at
MP2 and RPA levels converge dramatically slower than those of LDA, PBE and HSE06. The \textbf{\textit{k}}-mesh incompleteness 
error in C diamond remains to be 15 meV for both MP2 and RPA total energies with a dense $8\times8\times8$ $\boldsymbol{k}$-mesh.
% As shown above, the Monkhorst-Pack evenly sampled \textit{\textbf{k}}-meshes can provide a more efficient sampling for mean-field approximations.
We notice that other off-set evenly sampled \textit{\textbf{k}} meshes~\cite{chadi:1973A,marcus:1971}, 
the so-called mean-value point strategy~\cite{baldereschi:1973A}, and smearing sampling methods, 
such as the method of Methfessel and Paxton~\cite{methfessel:1989A}, the improved linear tetrahedron method~\cite{bloechl:1994A}, etc., 
have all been proposed to provide a better sampling with fewer (or even one) \textit{\textbf{k}} points, but at the semi-local DFT level of theory only.
It is interesting to investigate the influence of these sampling strategies on advanced correlation methods.

The computational expense of the canonical MP2 (and RPA) scales as $N_{\boldsymbol{k}}^3$ (and $N_{\boldsymbol{k}}^2$), where
$N_{\boldsymbol{k}}=n_{\boldsymbol{k}}^3$ is the total grid number in a given \textbf{\textit{k}} mesh. 
For C diamond and NAO-VCC-2Z basis set (with 28 basis functions per unit cell), the MP2 calculation with the $8\times8\times8$ $\boldsymbol{k}$-mesh
requires 10 hours to complete using 320 CPU cores of an Infiniband-connected Intel cluster with Intel Xeon E5-2680 v2 cores 
(2.8GHz, 20 cores per node). The periodic MP2 implementation in FHI-aims is a $\boldsymbol{k}$-mesh-oriented parallelization
in the framework of the Message Passing Interface (MPI), guaranteeing efficient MP2 calculations with thousands of cores. However,
with the optimistic assumption of a perfect parallel scalability, it would need over 2000 CPU cores to finish the MP2 calculation with a denser 
$\boldsymbol{k}$-mesh ($10\times10\times10$) in a comparable amount of time.
While the RPA calculations with such a dense $\boldsymbol{k}$-mesh and the NAO-VCC-2Z basis set
can be carried out at a reasonable cost, the remaining $\boldsymbol{k}$-mesh error is about 7 meV for C diamond.
The RPA calculations with denser $\boldsymbol{k}$-meshes and larger basis sets become infeasible as well.

A practical way to approach the complete \textbf{\textit{k}}-mesh limit (CKM) for advanced correlation methods is the two-point extrapolation 
in terms of an inverse relation between $\Delta E_{\textrm{total}}[n_{\textit{\textbf{k}}}]$ and $n_{\textit{\textbf{k}}}$:
\begin{equation}
 \label{Eq:k-mesh-relation}
 \begin{split}
     \Delta E_{\textrm{total}}[n_{\textit{\textbf{k}}}]&=A/n_{\textit{\textbf{k}}}^{\alpha}\\
     \log{\left(\Delta E_{\textrm{total}}[n_{\textit{\textbf{k}}}]\right)}&=\log\left(A\right)-\alpha\log{\left(n_{\textit{\textbf{k}}}\right)}
 \end{split}
\end{equation}
The exponential ``$\alpha$'' determines the speed of the \textbf{\textit{k}}-mesh convergence of different methods. 
On a logarithmic scale (see Eq.~\ref{Eq:k-mesh-relation} and Fig.~\ref{Fig:k_grid}), ``$\alpha$'' is the negative value of the slope of 
$\Delta E_{\textrm{total}}[n_{\textit{\textbf{k}}}]$ towards the CKM limit.
Previously, the exponential $\alpha=3.00$ has been successfully used to extrapolated MP2 and CCSD(T) cohesive energies for several materials~\cite{booth:2013A}.

\begin{table}
\begin{ruledtabular}
 \caption{Errors of extrapolated RPA and MP2 total energies per atom for five materials (in meV). 
	 The complete \textbf{\textit{k}}-mesh (CKM) extrapolation is performed from the different 
	 combinations of \textbf{\textit{k}} grids, namely CKM($n_{\boldsymbol{k}_1}$,$n_{\boldsymbol{k}_2}$), and with different exponentials $\alpha$. The errors listed are the deviations of
 E$_{\textrm{total}}$-E$_{\textrm{total}}^{\textrm{Ref}}$, where the reference E$_{\textrm{total}}^{\textrm{Ref}}$ for RPA is the extrapolated
 value from $8\times8\times8$ and $10\times10\times10$, CKM(8,10), with $\alpha=3.00$. For MP2, the reference is extrapolated by CKM(6,8) and $\alpha=3.00$.
 The mean absolute deviations (MADs) are shown for RPA and MP2 separately.
 }
 \label{Table:CKM}
 \begin{tabular}{ccccc}
%  & & ($n_{\boldsymbol{k}_1}$=4,$n_{\boldsymbol{k}_2}$=6)& ($n_{\boldsymbol{k}_1}$=4,$n_{\boldsymbol{k}_2}$=6) & ($n_{\boldsymbol{k}_1}$=6,$n_{\boldsymbol{k}_2}$=8)\\
        &Index&CKM(4,6)&CKM(4,6)&CKM(6,8)\\
        &in MSE & $\alpha=3.00$& $\alpha=3.95$ &$\alpha=3.00$\\
\hline
&&\multicolumn{3}{c}{RPA}\\
C	& 06 &	-13.8	&      -1.6	&	 0.4	\\
Si	& 07 &	-12.4	&	2.3	&	-1.3	\\
MgO	& 12 &	  1.5	&	2.4	&	 1.4	\\
BN	& 15 &	 -5.7	&	1.0	&	 1.3	\\
AlP	& 18 &	 -6.9	&	1.1	&	 1.1	\\
\hline
MAE	&    &	  8.0	&	1.6	&	 1.1	\\
\hline
&&\multicolumn{3}{c}{MP2}\\
C	& 06 &	-17.9	&      -1.9	&		\\
Si	& 07 &	-17.1	&	1.6	&		\\
MgO	& 12 &	  0.5	&	0.7	&		\\
BN	& 15 &	 -9.2	&	1.6	&		\\
AlP	& 18 &	 -9.5	&      -0.2	&		\\
\hline
MAD	&    &	 11.3	&	1.1	&		\\
 \end{tabular}
 \end{ruledtabular}
\end{table}

Table~\ref{Table:CKM} shows the performance of such \textbf{\textit{k}}-mesh extrapolation for
five selected materials (two elemental and three binary solids) 
from the MSE test set. The complete $\boldsymbol{k}$-mesh extrapolation from $8\times8\times8$ 
and $10\times10\times10$, CKM(8,10), are taken as 
the reference for the periodic RPA calculations. The deviations of the CKM(6,8) values are 
smaller than 1.5 meV for all five materials and the mean absolute deviation (MAD) is only 1.1 meV. 
Our results confirm the validity of the practical extrapolation 
(Eq.~\ref{Eq:k-mesh-relation}) with $\alpha=3.00$ for advanced methods provided that reasonable
dense $\boldsymbol{k}$-meshes are utilized. 
However, a notable decrease of the performance can be observed in the combination of CKM(4,6)
and $\alpha=3.00$. The MAD remains at about 8.0 meV (with the maximum error of 14 meV for C diamond) 
for the RPA total energies per atom. In this project, we optimize ``$\alpha$'' for the
CKM(4,6) extrapolation, and find that $\alpha=3.95$ improves the CKM(4,6) extrapolation 
consistently, resulting in the MAD of only 1.6 meV.

The CKM(6,8) values with $\alpha=3.00$ are chosen as the reference for MP2. Table \ref{Table:CKM} indicates that the CKM(4,6) extrapolation is not 
good enough when using the default choice of $\alpha=3.00$, and it can be effectively improved with an exponential value
$\alpha=3.95$ as optimized for RPA. Our results reveal that advanced correlation methods share a similar $\boldsymbol{k}$-mesh
convergence behavior, and thus the CKM(6,8) values with $\alpha=3.00$ are good enough to converge the error within 2.0 meV for insulators and semiconductors,
which is choosen for both MP2 and RPA calculations in this work (see Table \ref{Table:k-mesh}).

We notice that the periodic MP2 and CCSD(T) results reported in the literature
were often extrapolated using CKM(5,6) 
or even sparser CKM(3,4) with $\alpha=3.00$~\cite{booth:2013A}. It is an understandable compromise 
considering the computational expense with denser $\boldsymbol{k}$-meshes
and larger basis sets for these methods; However, in order to achieve $\boldsymbol{k}$-mesh-converged
values with the numerical uncertainty within 10 meV, an adapted exponential factor for CKM 
extrapolation is necessary.

\subsection{The basis set convergence}

The difficulty of using the GTO basis sets, the \emph{de facto} choice in quantum chemistry, 
to extended systems has been studied extensively in different kinds
of crystals mainly using the CRYSTAL program~\cite{gruneich:1998A,spackman:2001A,
champagne:2003A}. To quote the statement in their User's Manual (Chapter 10)~\cite{crystal09:2013A},
``as a rule, extended atomic basis sets, or 'triple zeta' type basis sets should be avoided (...),
because the outer functions are too diffuse.'' On the other hand, despite the predominate use
in calculations on periodic systems, the plane-wave basis set is ineffective to describe the 
localized core-electron states and usually needs to be used in conjunction with pseudopotentials. 

The NAO basis sets of FHI-aims\cite{scheffler:2009A,igor:2013A}, including \emph{tier-n} 
and NAO-VCC-\emph{n}Z, hold the promise to provide numerically well-converged total energies 
for mean-field approximations in \emph{all-electron} description,
as the minimal basis of both \emph{tier}-n and NAO-VCC-\emph{n}Z is composed of 
the exact core and valence orbitals of spherically symmetric free atoms. In the meantime, 
a confining potential can be used to generate the primitive NAOs 
that are exactly localized in a certain region, so that any numerical instability caused 
by an unnecessary overlap between the diffuse atom-centered basis functions can be 
significantly suppressed (see Refs.~\onlinecite{scheffler:2009A,igor:2013A} for details).

Table \ref{Table:tier-n} lists the basis set incompleteness errors in HSE06 total energies 
per atom using the \emph{tier}-$n$ series, which are the default choice for mean-field 
approximations in FHI-aims. The \emph{tier}-2 basis set is recommended, and \emph{tier}-3
is considered to be good enough to reach the complete basis set limit~\cite{scheffler:2009A}.
% The convergence accuracy of \emph{tier}-2 in the PBE total energies of symmetric dimers is about 10 meV~\cite{scheffler:2009A}. 
While the \emph{tier}-$n$ basis sets are optimized with respect to small molecules, i.e.\  symmetric dimers for all elements~\cite{scheffler:2009A}, our results 
demonstrate their transferability to the extended systems with diverse chemical environments. The average (maximum) error 
in \emph{tier}-2  and \emph{tier}-3  are about 8 meV (16 meV) and 2 meV (9 meV) in HSE06 total 
energies per atom of light elements' materials. Note that
the basis size of \emph{tier}-2 is slightly larger than Dunning's GTO triple-zeta basis set 
cc-pVTZ~\cite{dunning:1989A}, which has been recommended to
be avoided for calculations on solids~\cite{crystal09:2013A}. In our work, we do not encounter any self-consistent field (SCF) convergence problem 
with \emph{tier}-2 in all our calculations with GGAs and hybrid GGAs, including LDA, PBE, PBEsol, and HSE06. For meta-GGAs, the SCF convergence can be
achieved with \emph{tier}-2 as well, but some code-specific numerical parameters must be set carefully. For instance, we should use a finer numerical integration grid 
for the vdW systems~\cite{becke:2009A}, and it is necessary to introduce a threshold for the kinetic-energy related variable $\tau$, which removes values that do not effect the 
total energy but do cause sigularities when calculating the potential in the NAO framework~\cite{cremer:2007A}. Both parameters have been tested extensively, with
values chosen such that the errors these parameters introduce are in the sub-meV regime.

\begin{figure}
  \includegraphics[width=1.0\columnwidth]{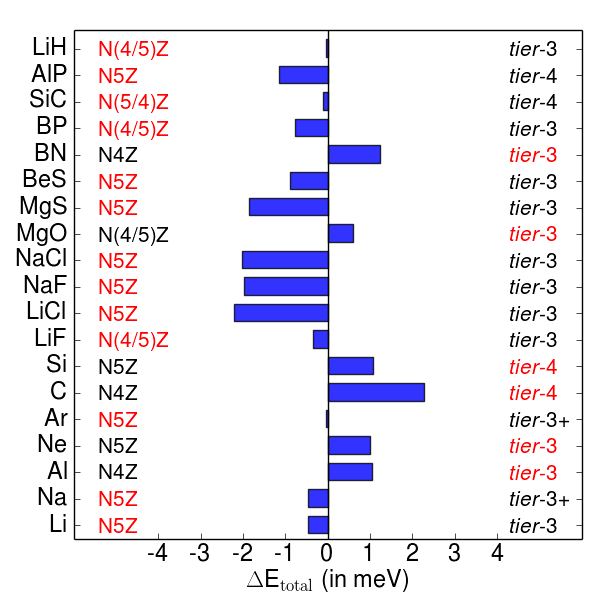}
  \caption{Energy differences of HSE06 total energies per atom with the best NAO-VCC-$n$Z (N$n$Z for short) and the best \emph{tier}-$n$
  ($\Delta$E$_{\rm total}$=E$_{\rm total}$[N$n$Z]-E$_{\rm total}$[\emph{tier}-$n$], in meV).
  The corresponding basis sets employed for each material are listed on the left and right hand side,
  respectively. \emph{tier}-3+ is the \emph{tier}-3 basis set plus 2 
  $s$-type and 2 $p$-type hydrogen-like basis functions from NAO-VCC-3Z. For binary crystals ``AB'', N($n_1$/$n_2$)Z is a short-hand notation of using NAO-VCC-$n_1$Z for A and 
  NAO-VCC-$n_2$Z for B. The basis sets in red deliver the lowest HSE06 results for each material, which are taken as the reference data in this work.}
  \label{Fig:basis_set}
\end{figure}

FHI-aims provides a larger \emph{tier} basis set (\emph{tier}-4) for some light elements.
%However, serious accidental ill-conditioning is arising with \emph{tier}-4, especially 
We notice that for the binaries with mainly ionic bonding characters (LiF - MgS), the
SCF procedure with tier-4 can fail.
%where the SCF procedure either converges to unphysical states or even breaks down. 
In this work, we report HSE06 total energies per atom using \emph{tier}-4 for C and Si with diamond structure, SiC and AlP
with zincblende structure. Compared with \emph{tier}-3 results, a noticeable change can be observed only for AlP (about 3 meV). It suggests that 
\emph{tier}-3 is competent to provide numerically well-converged HSE06 total energies for most 
of materials. Concerning the two worst cases of \emph{tier}-3,
i.e.\ Na bcc and Ar fcc, the basis set incompleteness errors are about 5 meV and 9 meV, respectively. 
For Na and Ar, the \emph{tier}-4 basis sets are unavailable. We then introduce 2 $s$-type and 2 $p$-type hydrogen-like functions from NAO-VCC-3Z as
an additional $s,p$ group to the \emph{tier}-3, denoted as \emph{tier}-3+. Table~\ref{Table:tier-n} and Fig.~\ref{Fig:basis_set} reveal that 
such $s,p$ group effectively compensates for the basis set incompleteness in \emph{tier}-3, resulting in the HSE06 total energies per atom with CBS quality
for Na bcc and Ar fcc.

\begin{table}
\begin{ruledtabular}
 \caption{Basis set errors in HSE06 total energies per atom (in meV). The reference data are the lowest energies in two sequences 
 of NAO basis sets, i.e.\ \emph{tier-n} and NAO-VCC-$n$Z.}
 \label{Table:tier-n}
 \begin{tabular}{ccccc|cccc}
	&\multicolumn{4}{c|}{\emph{tier-n}}&\multicolumn{4}{c}{NAO-VCC-$n$Z}\\
	&$n$=1	&$n$=2	&$n$=3	&$n$=4 & $n$=2 & $n$=3 & $n$=4 & $n$=5\\
\hline
Li	&	17	&	4	&	0	&	--	&	55	&	18	&	1	&	0	\\
Na	&	32	&	16	&	5	&	0\footnotemark[1]	&	22	&	11	&	3	&	0	\\
Al	&	16	&	4	&	0	&	--	&	50	&	13	&	1	&	--	\\
Ne	&	13	&	11	&	0	&	--	&	16	&	6	&	4	&	1	\\
Ar	&	16	&	11	&	9	&	0\footnotemark[1]	&	16	&	7	&	2	&	0	\\
C	&	46	&	2	&	0	&	0	&	67	&	23	&	2	&	--	\\
Si	&	32	&	10	&	1	&	0	&	66	&	22	&	4	&	1	\\
LiF	&	33	&	2	&	0	&	--	&	44	&	16	&	0	&	0\footnotemark[2]	\\
LiCl	&	23	&	7	&	2	&	--	&	42	&	15	&	2	&	0	\\
NaF	&	23	&	7	&	2	&	--	&	44	&	14	&	2	&	0	\\
NaCl	&	21	&	10	&	2	&	--	&	32	&	12	&	3	&	0	\\
MgO	&	61	&	11	&	0	&	--	&	87	&	18	&	2	&	1\footnotemark[2]	\\
MgS	&	55	&	13	&	2	&	--	&	69	&	18	&	4	&	0	\\
BeS	&	47	&	6	&	1	&	--	&	113	&	38	&	4	&	0	\\
BN	&	53	&	2	&	0	&	--	&	115	&	17	&	1	&	--	\\
BP	&	19	&	5	&	1	&	--	&	79	&	22	&	3	&	0\footnotemark[2]	\\
SiC	&	54	&	9	&	1	&	0	&	130	&	29	&	6	&	0\footnotemark[2]	\\
AlP	&	23	&	9	&	4	&	1	&	87	&	22	&	5	&	0	\\
LiH     &       11	&	1	&	0	&	--	&	48	&	11	&	0	&	0\footnotemark[2]	\\
\hline
Average	&	31	&	8	&	2	&		&	62	&	18	&	3	&		\\
 \end{tabular}
 \end{ruledtabular}
 \footnotetext[1]{HSE06 results of Na bcc and Ar fcc are calculated using \emph{tier}-3+ which is \emph{tier}-3 plus 2 $s$-type and 2 $p$-type
 basis functions from NAO-VCC-3Z.}
 \footnotetext[2]{For a binary crystal AB, a hybrid basis set with NAO-VCC-$n_1$Z for A and NAO-VCC-$n_2$Z for B is denoted as N($n_1$/$n_2$)Z.
 These are N(4/5)Z for LiF, MgO and LiH rocksalts, BP zincblende, and N(5/4)Z for SiC.}
\end{table}

The NAO basis functions used in the \emph{tier-n} series are apt to saturate the bonding region in the middle of two atoms, as they
are optimized to minimize the LDA total energies of symmetric dimers~\cite{scheffler:2009A}. In contrast, the sequence of NAO-VCC-$n$Z
basis sets is determined by minimizing the RPA total energies of atoms~\cite{igor:2013A}, thus introducing more compact NAO basis functions 
than the \emph{tier}-$n$ series. With similar basis sizes, NAO-VCC-3Z delivers a larger basis set incompleteness error (18 meV on average)
than \emph{tier}-2. It indicates the necessity of introducing diffuse atom-centered basis functions
in small basis sets to balance the aforementioned incompleteness in core and bonding regions,
which renders a better performance for mean-field approximations. However, such discrepancy can 
be reduced with the increase of the index ``$n$'' in both sequences, the average
error being only 2 meV and 3 meV for \emph{tier}-3 and NAO-VCC-4Z with similar basis sizes. 

NAO-VCC-5Z is the largest NAO basis set in FHI-aims, and yields the lowest HSE06 total energies 
per atom for most of the materials in the test set. 
However, SCF convergence problem occurs at Al fcc, C diamond, LiF and MgO with rocksalt structure, 
BN, BP and SiC with zincblende structure. For binary crystals ``AB'', we tried a hybrid basis 
set strategy N($n_1$/$n_2$)Z, which is a short-hand notation of
using NAO-VCC-$n_1$Z for ``A'' and NAO-VCC-$n_2$Z for ``B''. 
Compared with NAO-VCC-4Z results, better HSE06 total energies per atom can be obtained 
without SCF convergence problem for LiF, MgO, BP and LiH with N(4/5)Z and for SiC with N(5/4)Z.

Fig.~\ref{Fig:basis_set} presents the differences of the HSE06 total energies per atom ($\Delta E_{\rm total}$) between the
best NAO-VCC-$n$Z and \emph{tier}-$n$ basis sets. The discrepancy $\Delta E_{\rm total}$ is very small, about 1 meV on average.
Considering that NAO-VCC-$n$Z and \emph{tier}-n are designed for completely different purposes, such an agreement with each other
clearly indicates that the complete basis set limit has been approached for the calculations of HSE06 total energies per atom in \textit{all-electron}
description.

Other (semi-)local DFAs share a similar basis set convergence as HSE06. The observation and discussion here are then transferable.
In addition to HSE06, one can find a comprehensive basis set convergence test with LDA, PBE, TPSS, M06-L, and SCAN in the MSE web site. 
For PBEsol, the best NAO basis sets determined by PBE are used to calculate the reference data directly.
% to LDA and PBE. However, the SCF convergence with \emph{tier}-4 and/or even \emph{tier}-3 becomes more difficult to achieve, especially for the binaries 
% with mainly ionic bonding characters (LiF - MgS). The basis sets used to generate the reference data for LDA and PBE will be presented and discussed briefly 
% in the appendix.

For MP2 and RPA calculations, the results are extrapolated to the CBS limit using a 
two-point extrapolation formula~\cite{klopper:1997A,igor:2013A} based on
NAO-VCC-3Z and NAO-VCC-4Z, namely CBS(3,4): 
\begin{equation}
	\label{Eq:CBS}
	E[\infty]=\frac{E[3]3^3-E[4]4^3}{3^3-4^3}
\end{equation}
Here $E[n]$ is the RPA or MP2 total energy per atom using NAO-VCC-$n$Z basis sets
(with $n$=3 and 4). The accuracy of the CBS(3,4) scheme has been demonstrated
in the RPA and MP2 calculations of 18 symmetric dimers from H$_2$ to Ar$_2$~\cite{igor:2013A}. Taking the CBS extrapolation values from aug-cc-pV5Z
and 6Z as the reference, the average basis set error in CBS(3,4) values is about 13 meV. However, for solids, it is impossible to
perform the RPA and MP2 calculations using very large and diffuse GTO basis sets, i.e.\ aug-cc-pV5Z and 6Z. For LDA, PBE and HSE06, we observe a 
very good transferability of \emph{tier}-$n$ and NAO-VCC-$n$Z from isolated molecules to extended solids as shown in Table~\ref{Table:tier-n}. 
We argue that the performance of the CBS(3,4) strategy in molecules is also transferable to solids, which can be used to converge the RPA and 
MP2 total energies per atom in the MSE test set with a similar numerical uncertainty on average. 
While the extrapolation from NAO-VCC-4Z and NAO-VCC-5Z shall further reduce the numerical uncertainty of the reference data~\cite{igor:2013A}, NAO-VCC-5Z is too expensive to be used
for periodic MP2 and RPA calculations at present. 

The $\boldsymbol{k}$-mesh convergence of MP2 and RPA methods tested in the previous section is carried out together with the smallest NAO-VCC-2Z 
basis sets (see Fig.~\ref{Fig:k_grid}). However, it would be impossible to perform MP2 calculations with $8\times8\times8$ for NAO-VCC-4Z, and 
sometimes even a NAO-VCC-3Z calculation is too expensive, because of the unfavorable scaling of 
the MP2 method with respect to the number of $\boldsymbol{k}$ points.

\begin{table}
\begin{ruledtabular}
 \caption{Errors in the MP2 total energies per atom using different basis sets for C diamond (in meV). 
 The reference is the CKM(6,8) extrapolated values with $\alpha=3.00$ (see Eq.~\ref{Eq:k-mesh-relation}).}
 \label{Table:k_mesh_basis_set}
 \begin{tabular}{cccc}
$\boldsymbol{k}$-mesh	&	NAO-VCC-2Z	&	NAO-VCC-3Z	&	NAO-VCC-4Z	\\
\hline
($4\times4\times4$)	&	117	&	114	&	115	\\
($6\times6\times6$)	&	22	&	22	&	22	\\
($8\times8\times8$)	&	9	&	9	&	9\footnotemark[1]	\\
 \end{tabular}
 \footnotetext[1]{For NAO-VCC-4Z, the MP2 total energy per atom with the $8\times8\times8$ $\boldsymbol{k}$-mesh is extrapolated using the
 energy difference between $\boldsymbol{k}$-grids $6\times6\times6$ and $8\times8\times8$ with the NAO-VCC-3Z basis set.}
 \end{ruledtabular}
\end{table}

Taking C diamond as an example, Table~\ref{Table:k_mesh_basis_set} shows the $\boldsymbol{k}$-mesh convergence of the MP2 total energies per atom for different
basis sets. The most time consuming result in this table is the combination of NAO-VCC-3Z and 
the $8\times8\times8$ $\boldsymbol{k}$-mesh, which requires about 4 days using 320 CPU cores of an 
Infiniband-connected Intel cluster with Intel Xeon E5-2680 v2 cores (2.8 GHz, 20 cores per node). As illustrated in Table~\ref{Table:k_mesh_basis_set}, 
while the $\boldsymbol{k}$-mesh error of a $4\times4\times4$ mesh varies with the basis set employed by several meV, it becomes almost independent of the
choice of basis sets for $6\times6\times6$ and $8\times8\times8$ meshes. Ohnishi \emph{et al.}~\cite{hirata:2010A} and Gr\"uneis \emph{et al.}~\cite{gruneis:2011A}
had a similar observation in MP2 and CCSD calculations using the GTO-type and plane-wave basis sets, respectively. In consequence, they concluded that the 
long-range behavior of the correlation energy depends mostly on the low-lying excitations, and proposed their progressive downsampling technique to approach
the complete $\boldsymbol{k}$-mesh and basis set limit for advanced correlation methods.

In this work, we adopt a similar downsampling concept to estimate the MP2 total energies per atom with the NAO-VCC-4Z basis set and the $8\times8\times8$
\textbf{\textit{k}}-grid:
$E_{\rm total}^{\textrm{4Z}}[n_{\boldsymbol{k}}=8]$
\begin{equation}
   \label{Eq:downsampling}
   E_{\rm total}^{\textrm{4Z}}[n_{\boldsymbol{k}}=8] = E_{\rm total}^{\textrm{4Z}}[n_{\boldsymbol{k}}=6] + \Delta E_{\rm total}^{n\textrm{Z}}[n_{\boldsymbol{k}}=8]
\end{equation}
using the energy change between $6\times6\times6$ and $8\times8\times8$ with the NAO-VCC-3Z (or 2Z) basis set 
$\Delta E_{\rm total}^{n\textrm{Z}}[n_{\boldsymbol{k}}=8]$ ($n=3$ or 2) of:
\begin{equation}
   \Delta E_{\rm total}^{n\textrm{Z}}[n_{\boldsymbol{k}}=8] = E_{\rm total}^{n\textrm{Z}}[n_{\boldsymbol{k}}=8] - E_{\rm total}^{n\textrm{Z}}[n_{\boldsymbol{k}}=6]
\end{equation}
In summary, the reference data for the MP2 and RPA total energies (and also valence correlation energies) per atom are obtained by the combination
strategy of CKM(6,8), CBS(3,4) and $\Delta E_{\rm total}^{\textrm{nZ}}[n_{\boldsymbol{k}}=8]$ with ($n=3$ or 2). The error bar is estimated to be 20 meV
for these MP2 and RPA reference data. The numerical uncertainty in the combination strategy is dominated by the CBS(3,4) extrapolation,
i.e.\ 15 meV for CBS(3,4), 2 meV for CKM(6,8), and less than 1 meV for $\Delta E_{\rm total}^{\textrm{3Z}}[n_{\boldsymbol{k}}=8]$
or $\Delta E_{\rm total}^{\textrm{2Z}}[n_{\boldsymbol{k}}=8]$, respectively.

\subsection{Lattice constant, bulk modulus, and cohesive energy}
Based on the efforts discussed above to achieve the complete $\boldsymbol{k}$-mesh and basis set convergence in terms of absolute total energy for various levels of theory, 
we next focus on the calculated lattice constants, bulk moduli and cohesive energies using these parameters.

To find the optimized lattice parameters for each material and each method, we calculate seven points within a range of $\pm5$~\% around the initial value of the lattice 
constant, and fit the respective (volume, energy) data points to the Birch-Murnaghan equation of state to obtain the optimized lattice constant. If all 7 lattice constants used
to generate the equation of state lie within a range of $\pm7$~\% around the optimized value, the value is taken as the final, optimized lattice constant $a_0$. 
Otherwise, this optimized lattice constant is used as a new starting point and more (volume, energy) data points are calculated, so that finally all materials' optimized lattice constants are obtained
from an equation of state fitted to seven points with lattice constant values in a range of $\pm$5\% to $\pm$7\% around the final, optimized lattice constant $a_0$. We 
have carefully tested the stability of this procedure using the PBE functional. The obtained optimized lattice constants and bulk moduli are accurate within 0.02~{\AA} 
and 0.1~GPa, respectively. 

The cohesive energies $E^M_{coh}$ of the materials are then calculated at the optimized lattice constant $a_0$:
\begin{equation}
	\label{Eq:E_coh}
	E^M_{coh} =E^M / {N_{atom}}- {\sum_{atom}} E^{atom} / {N_{atom}}
\end{equation}
where $N_{atom}$ is the number of atoms in the unit cell, $E^M$ the total energy of the unit cell 
for material \emph{M}, and the sum is taken over the total energies $E^{atom}$ of the constituent
atoms in their spin-polarized symmetry-broken ground state, i.e. no fractional occupancies.

For (semi-)local and hybrid DFAs, the best basis sets and $\boldsymbol{k}$-grid settings recommended in Table~\ref{Table:k-mesh} and Fig.~\ref{Fig:basis_set}
are used to produce numerically well-converged lattice constants, bulk moduli and cohesive energies. These reference data are in \textit{all-electron} description. 
The post-processing MP2 and RPA correlations are frozen-core, but evaluated based on the Hartree-Fock, PBE and/or PBE0 orbitals in \textit{all-electron} description.
The basis set convergence of
these properties are well-documented for (semi-)local and hybrid DFAs. However, the slow basis set convergence together with an unfavorable
computational scaling makes it a big challenge to perform a systematic investigation of the basis set convergence of these materials' properties at
MP2 and RPA levels. We present here the MP2 results of five selected materials from the MSE test sets in Table~\ref{Table:MP2-results}. 
To the best of our knowledge, it is the first report of the basis set convergence of MP2 for condensed matter systems using NAO basis sets.
The readers can easily access the MP2 and RPA data on convergence test for all 13 materials 
in our MSE web site.

Table~\ref{Table:MP2-results} reveals that the basis set convergence is relatively fast 
in the MP2 calculations of lattice constants.
NAO-VCC-3Z is good enough to provide a well-converged lattice constant (the convergence
criterion is 0.01 \AA) for all five materials (for details see also the first example 
in Sec~\ref{Sec:website}). In agreement with the previous investigations using plane-wave 
basis sets~\cite{gruneis:2010A,booth:2013A} or a Gaussian and 
plane waves hybrid approach~\cite{mauro:2013A}, the slow basis set convergence of MP2 cohesive energies is observed with NAO basis sets as well.
Compared with CBS(3,4) values, the basis set incompleteness errors at NAO-VCC-4Z still remain about 60-130 meV. 
Such slow convergence arises from the slower basis set convergence of the MP2 total energies in bulks than that in free atoms.
It is well-documented in quantum chemistry that the accurate geometry information for the MP2 and other
advanced correlation methods can be obtained with the basis sets of triple-zeta quality~\cite{wiberg:2004A}, but the converged 
atomization energy or other energy differences cannot be achieved with finite basis sets~\cite{curtiss:1989A,morgan:1992A,feller:2000A}. 
Our results confirm that this conclusion is also valid for solids. 
%The basis set convergence of the energetic second-order response properties is a significant challenge in both 
%quantum chemistry and materials science. 
While the bulk moduli obtained at the NAO-VCC-4Z basis set are not well-converged, the basis set error remains about 3 -- 7 GPa.

Inspecting Table~\ref{Table:MP2-results} also reveals the capability of NAO-VCC-$n$Z to 
provide a consistently improvable 
description of cohesive properties. In general, the calculated MP2 cohesive energies
increase with the cardinal index ($n$) of NAO-VCC-$n$Z. On the one hand, 
it allows for an accurate extrapolation to the CBS limit; and on the other hand, the 
NAO-VCC-4Z values set up a quite rigorous lower bound for the converged values of MP2. 
In other words, any results smaller than the NAO-VCC-4Z values might be unreliable.

\begin{table}
\begin{ruledtabular}
 \caption{MP2 cohesive energies per atom, lattice constants, and bulk moduli. All results have been
 extrapolated to the complete $\boldsymbol{k}$-mesh limit using CKM(6,8).
 The VASP results are listed as well.}
 \label{Table:MP2-results}
 \begin{tabular}{ccccc}
                     &  Basis set  &  $E_{coh}$(eV)  &  $a_0$(\AA)  &  $B_0$(GPa)\\
                  \hline
                  C  &  NAO-VCC-2Z  &  7.65  &  3.56  &  451 \\
                     &  NAO-VCC-3Z  &  7.81  &  3.55  &  454 \\
                     &  NAO-VCC-4Z  &  7.96  &  3.54  &  454 \\
                     &  CBS(3,4)  &  8.08  &  3.54  &  454 \\
                     &  VASP\footnotemark[1]  &  7.97  &  3.55  &  450 \\
                     &  VASP\footnotemark[2]  &  8.04  &    &    \\
                  \hline
                  Si &  NAO-VCC-2Z  &  4.62  &  5.44  &  98 \\
                     &  NAO-VCC-3Z  &  4.92  &  5.41  &  100 \\
                     &  NAO-VCC-4Z  &  5.07  &  5.41  &  100 \\
                     &  CBS(3,4)  &  5.21  &  5.40  &  100 \\
                     &  VASP\footnotemark[1]  & 5.05  &  5.42  &  100 \\
                     &  VASP\footnotemark[2]  &     &    &   \\
                  \hline
                   BN  &  NAO-VCC-2Z  &  6.84  &  3.62  &  387 \\
                     &  NAO-VCC-3Z  &  7.01  &  3.59  &  392 \\
                     &  NAO-VCC-4Z  &  7.14  &  3.59  &  375 \\
                     &  CBS(3,4)  &  7.25  &  3.58  &  368 \\
                     &  VASP\footnotemark[1]  &  7.12  &  3.61  &  395 \\
                     &  VASP\footnotemark[2]  &  7.15  &     &    \\
                  \hline
                  MgO  &  NAO-VCC-2Z  &  5.11  &  4.23  &  160 \\
                     &  NAO-VCC-3Z  &  5.39  &  4.23  &  156 \\
                     &  NAO-VCC-4Z  &  5.47  &  4.24  &  156 \\
                     &  CBS(3,4)  &  5.53  &  4.24  &  156 \\
                     &  VASP\footnotemark[1]  &  5.35  &  4.23  &  153 \\
                     &  VASP\footnotemark[2]  &     &     &    \\

                  \hline
                  AlP  &  NAO-VCC-2Z  &  4.06  &  5.48  &  92 \\
                     &  NAO-VCC-3Z  &  4.41  &  5.45  &  94 \\
                     &  NAO-VCC-4Z  &  4.55  &  5.45  &  95 \\
                     &  CBS(3,4)  &  4.67  &  5.45  &  95 \\
                     &  VASP\footnotemark[1] &  4.32  &  5.46  &  93 \\
                     &  VASP\footnotemark[2]  &   4.63  &     &   \\
 \end{tabular}
 \footnotetext[1]{In \cite{gruneis:2010A}, the MP2 results with VASP were extrapolated to the CBS limit. The \textit{\textbf{k}}-mesh was sampled
	by a composite scheme. The Hatree-Fock orbitals were generated in the pseudo-potential framework.}
\footnotetext[2]{In \cite{booth:2013A}, the MP2 results with VASP were extrapolated by CKM(5,6) with $\alpha=3$. 
        The Hartree-Fock orbitals were generated in the pseudo-potential framework.}
 \end{ruledtabular}
\end{table}

The canonical MP2 method was implemented in VASP for periodic systems in 2009 \cite{gruneis:2009A}.
Later, Gr\"uneis \textit{et al.} calculated the MP2 cohesive properties of 13 materials in 2010 \cite{gruneis:2010A}.
To approach the complete basis set limit in the framework of the projector-augmented-wave (PAW) method using plane-wave basis sets,
they proposed an inverse relation between the MP2 correlation energy and the energy cutoff $E_{\chi}$ that represents the overlap
charge densities for the CBS extrapolation. 
Meanwhile, an efficient composite scheme was proposed to reduce the computational cost due to the unfavorable scaling of the $\boldsymbol{k}$-point number.
To be specific, the Hartree-Fock total energy is calculated with a dense $10\times10\times10$ $\Gamma$-centered
\textit{\textbf{k}}-mesh, but the MP2 second-order direct and exchange terms are evaluated with $6\times6\times6$ and $3\times3\times3$ 
$\boldsymbol{k}$-meshes, respectively.
In 2013,  Booth \textit{et al.} updated the MP2 cohesive energies of three materials \cite{booth:2013A}, including C diamond, BN and AlP with zincblende structure.
The main difference is to replace the composite scheme by an extrapolation scheme from $5\times5\times5$ and $6\times6\times6$, i.e. CKM(5,6) with $\alpha=3$,
to approach the converged \textit{\textbf{k}}-mesh sampling. Table \ref{Table:MP2-results} also lists the canonical MP2 cohesive properties calculated 
with VASP. 
FHI-aims predicts very similar cohesive energies to the VASP values with $\boldsymbol{k}$-grid extrapolation scheme. The difference is about 40 meV for C diamond
and AlP zincblende, and 100 meV for BN in zincblende.
For MgO and Si, the discrepancy of over 170 meV between FHI-aims and VASP values shall be 
ascribed to the $\boldsymbol{k}$-grid incompleteness of the composite scheme
used in the previous VASP calculations~\cite{gruneis:2010A}.

\section{Conclusions}
\label{Sec:conclusions}
With the materials science and engineering (MSE) test set, we have accomplished the first step 
towards a representative test set with well-defined and relevant cohesive and electronic 
properties and reference values 
obtained from a hierarchy of the first-principles calculations. The accuracy of the reference values 
in the MSE test set is mainly determined
by the well-defined numerical setting of each applied DFA. A strong effort has been made to 
provide numerically converged values from state-of-the-art theory, including periodic
MP2, and RPA calculations in the complete basis set limit and the complete 
\textbf{\textit{k}}-space limit as well. At present, we provide 14 accurate data for MP2 and RPA. 
%Feasibility of rigorous numerical convergence tests for these high-level methods has been achieved
%by taking advantage of the FHI-aims all-electron approach, utilizing correlation-consistent
%numerical atom-centered orbital basis sets, using efficient techniques 
%dealing with four-center integrations, and applying advanced parallelization strategies. 

A web site of the MSE test set, \href{http://mse.fhi-berlin.mpg.de}{\texttt{http://mse.fhi-berlin.mpg.de}},
is equipped with a multi-mode access framework, versatile visualization, and a linear
regression tool for post processing data analysis.
In this work, we demonstrate that these features dramatically assist the post-processing data analysis that is necessary to detect the numerical error in calculations
and uncover the intrinsic limitations of DFAs.
The presented paradigm for the test set construction is applicable to any new material 
and materials' property.
% With respect to the assessment of numerical errors in advanced correlation methods for materials science, 

% heavy elements and non-cubic structures, defects, and surface should be included in future work.

\section{Acknowledgments}
The authors thank Fawzi Mohamed for his support in setting up the MSE web page, 
Steffen Kangowski for creating the layout of the MSE web page, and Norina A. Richter and Toktam
Morshedloo for their contribution in the preparation stage of the MSE project.
AJL is grateful to the Royal Society of Chemistry for the award of a Researcher Mobility 
Fellowship, and acknowledges the use of the ARCHER high performance computing facilities,
and associated support services, via membership of the UK HPC Materials 
Chemistry Consortium (EP/L000202).

\bibliography{paper}% Produces the bibliography via BibTeX.

\begin{thebibliography}{144}
\providecommand{\natexlab}[1]{#1}
\providecommand{\url}[1]{\texttt{#1}}
\expandafter\ifx\csname urlstyle\endcsname\relax
  \providecommand{\doi}[1]{doi: #1}\else
  \providecommand{\doi}{doi: \begingroup \urlstyle{rm}\Url}\fi

\bibitem[Curtiss et~al.(2005)Curtiss, Redfern, and Raghavachari]{curtiss:2005A}
Larry~A. Curtiss, Paul~C. Redfern, and Krishnan Raghavachari.
\newblock \emph{J. Chem. Phys.}, 123:\penalty0 124107--12, 2005.

\bibitem[Schultz et~al.(2005{\natexlab{a}})Schultz, Zhao, and
  Truhlar]{truhlar:2005A}
N.~E. Schultz, Y.~Zhao, and D.~G. Truhlar.
\newblock \emph{J. Phys. Chem. A}, 109:\penalty0 4388--4403,
  2005{\natexlab{a}}.

\bibitem[Schultz et~al.(2005{\natexlab{b}})Schultz, Zhao, and
  Truhlar]{truhlar:2005B}
N.~E. Schultz, Y.~Zhao, and D.~G. Truhlar.
\newblock \emph{J. Phys. Chem. A}, 109:\penalty0 11127--11143,
  2005{\natexlab{b}}.

\bibitem[Zhao et~al.(2005)Zhao, Gonz{\' a}lez-Garc{\' i}a, and
  Truhlar]{truhlar:2005C}
Y.~Zhao, N.~Gonz{\' a}lez-Garc{\' i}a, and D.~G. Truhlar.
\newblock \emph{J. Phys. Chem. A}, 109:\penalty0 2012--2018, 2005.

\bibitem[Zhao and Truhlar(2005)]{truhlar:2005D}
Yan Zhao and Donald~G. Truhlar.
\newblock \emph{J. Chem. Theory Comput.}, 1:\penalty0 415--432, 2005.

\bibitem[Luo et~al.(2014)Luo, Averkiev, Yang, Xu, and Truhlar]{truhlar:2014A}
Sijie Luo, Boris Averkiev, Ke~R. Yang, Xuefei Xu, and Donald~G. Truhlar.
\newblock \emph{J. Chem. Theory Comput.}, 10:\penalty0 102--121, 2014.

\bibitem[Jure{\v c}ka et~al.(2006)Jure{\v c}ka, {\v S}poner, {\v C}ern{\' y},
  and Hobza]{hobza:2006A}
P~Jure{\v c}ka, J~{\v S}poner, J~{\v C}ern{\' y}, and P~Hobza.
\newblock \emph{Phys. Chem. Chem. Phys.}, 8:\penalty0 1985--1993, 2006.

\bibitem[R{\v e}z{\' a}{\v c} et~al.(2011)R{\v e}z{\' a}{\v c}, Riley, and
  Hobza]{hobza:2011A}
Jan R{\v e}z{\' a}{\v c}, Kevin~E. Riley, and Pavel Hobza.
\newblock \emph{J. Chem. Theory Comput.}, 7:\penalty0 2427--2438, 2011.

\bibitem[Karton et~al.(2008)Karton, Tarnopolsky, Lam{\` e}re, Schatz, and
  Martin]{martin:2008A}
Amir Karton, Alex Tarnopolsky, Jean-François Lam{\` e}re, George~C. Schatz,
  and Jan M.~L. Martin.
\newblock \emph{J. Phys. Chem. A}, 112:\penalty0 12868--12886, 2008.

\bibitem[Bryantsev et~al.(2009)Bryantsev, Diallo, van Duin, and
  Goddard]{goddard:2009A}
Vyacheslav~S. Bryantsev, Mamadou~S. Diallo, Adri C.~T. van Duin, and William~A.
  Goddard.
\newblock \emph{J. Chem. Theory Comput.}, 5:\penalty0 1016--1026, 2009.

\bibitem[Jiang et~al.(2012)Jiang, DeYonker, Determan, and Wilson]{jiang:2012A}
Wanyi Jiang, Nathan~J. DeYonker, John~J. Determan, and Angela~K. Wilson.
\newblock \emph{J. Phys. Chem. A}, 116:\penalty0 870--885, 2012.

\bibitem[Wu and Xu(2008)]{xu:2008A}
J.~M. Wu and X.~Xu.
\newblock \emph{J. Chem. Phys.}, 129:\penalty0 164103--11, 2008.

\bibitem[Wu and Xu(2009)]{xu:2009A}
J.~M. Wu and X.~Xu.
\newblock \emph{J. Comput. Chem.}, 30:\penalty0 1424--1444, 2009.

\bibitem[Goerigk and Grimme(2010)]{grimme:2010A}
Lars Goerigk and Stefan Grimme.
\newblock \emph{J. Chem. Theory Comput.}, 6:\penalty0 107--126, 2010.

\bibitem[Goerigk and Grimme(2011)]{grimme:2011A}
Lars Goerigk and Stefan Grimme.
\newblock \emph{J. Chem. Theory Comput.}, 7:\penalty0 291--309, 2011.

\bibitem[Jung et~al.(2005)Jung, Sodt, Gill, and Head-Gordon]{head-gordon:2005A}
Yousung Jung, Alex Sodt, Peter M.~W. Gill, and Martin Head-Gordon.
\newblock \emph{Proc. Natl. Acad. Sci. {USA}}, 102:\penalty0 6692 --6697, 2005.

\bibitem[Hohenstein and Sherrill(2010)]{sherrill:2010A}
Edward~G. Hohenstein and C.~David Sherrill.
\newblock \emph{J. Chem. Phys.}, 133:\penalty0 104107, 2010.

\bibitem[Ren et~al.(2012{\natexlab{a}})Ren, Rinke, Blum, Wieferink, Tkatchenko,
  Sanfilippo, Reuter, and Scheffler]{ren:2012A}
Xinguo Ren, Patrick Rinke, Volker Blum, Juergen Wieferink, Alexandre
  Tkatchenko, Andrea Sanfilippo, Karsten Reuter, and Matthias Scheffler.
\newblock \emph{New. J. Phys.}, 14:\penalty0 053020--60, 2012{\natexlab{a}}.

\bibitem[Mahler and Wilson(2013)]{mahler:2013A}
Andrew Mahler and Angela~K. Wilson.
\newblock \emph{J. Chem. Theory Comput.}, 9:\penalty0 1402--1407, 2013.

\bibitem[Zhang et~al.(2010{\natexlab{a}})Zhang, Luo, and Xu]{xu:2010A}
IY~Zhang, Y~Luo, and X~Xu.
\newblock \emph{J. Chem. Phys.}, 133:\penalty0 104105--12, 2010{\natexlab{a}}.

\bibitem[Papajak and Truhlar(2012)]{truhlar:2012A}
Ewa Papajak and Donald~G. Truhlar.
\newblock \emph{J. Chem. Phys.}, 137, 2012.

\bibitem[Goldey and Head-Gordon(2012)]{head-gordon:2012A}
Matthew Goldey and Martin Head-Gordon.
\newblock \emph{J. Phys. Chem. Lett.}, 3:\penalty0 3592--3598, 2012.

\bibitem[Eshuis and Furche(2012)]{furche:2012A}
Henk Eshuis and Filipp Furche.
\newblock \emph{J. Chem. Phys.}, 136:\penalty0 084105, 2012.

\bibitem[Zhang et~al.(2013)Zhang, Ren, Rinke, Blum, and Scheffler]{igor:2013A}
Igor~Ying Zhang, Xinguo Ren, Patrick Rinke, Volker Blum, and Matthias
  Scheffler.
\newblock \emph{New J. Phys.}, 15:\penalty0 123033, 2013.

\bibitem[Curtiss et~al.(1997)Curtiss, Raghavachari, Redfern, and
  Pople]{curtiss:1997A}
L.~A. Curtiss, K~Raghavachari, P.~C. Redfern, and J.~A. Pople.
\newblock \emph{J. Chem. Phys.}, 106:\penalty0 1063--1079, 1997.

\bibitem[Curtiss et~al.(2000)Curtiss, Raghavachari, Redfern, and
  Pople]{curtiss:2000A}
L.~A. Curtiss, K~Raghavachari, P.~C. Redfern, and J.~A. Pople.
\newblock \emph{J. Chem. Phys.}, 112:\penalty0 7374--7383, 2000.

\bibitem[Xu and Goddard(2004{\natexlab{a}})]{xu:2004A}
X.~Xu and W.~A. Goddard.
\newblock \emph{Proc. Natl. Acad. Sci. {USA}}, 101:\penalty0 2673--2677,
  2004{\natexlab{a}}.

\bibitem[Xu and Goddard(2004{\natexlab{b}})]{xu:2004B}
X.~Xu and W.~A. Goddard.
\newblock \emph{J. Chem. Phys.}, 121:\penalty0 4068--4082, 2004{\natexlab{b}}.

\bibitem[Grimme(2006)]{grimme:2006A}
Stefan Grimme.
\newblock \emph{J. Chem. Phys.}, 124:\penalty0 034108--16, 2006.

\bibitem[Wu and Xu(2007)]{xu:2007A}
J.~M. Wu and X.~Xu.
\newblock \emph{J. Chem. Phys.}, 127:\penalty0 214105--214113, 2007.

\bibitem[Zhang et~al.(2009)Zhang, Xu, and Goddard]{xu:2009B}
Y~Zhang, X~Xu, and W.~A. Goddard.
\newblock \emph{Proc. Natl. Acad. Sci. {USA}}, 106:\penalty0 4963--4968, 2009.

\bibitem[Zhang et~al.(2011)Zhang, Xu, Jung, and Goddard]{xu:2011A}
Igor~Ying Zhang, Xin Xu, Yousung Jung, and W.~A. Goddard.
\newblock \emph{Proc. Natl. Acad. Sci. {USA}}, 108:\penalty0 19896--19900,
  2011.

\bibitem[Chai and Head-Gordon(2009)]{head-gordon:2009A}
J.~D. Chai and M~Head-Gordon.
\newblock \emph{J. Chem. Phys.}, 131:\penalty0 174105--13, 2009.

\bibitem[Ren et~al.(2011)Ren, Tkatchenko, Rinke, and Scheffler]{ren:2011A}
Xinguo Ren, Alexandre Tkatchenko, Patrick Rinke, and Matthias Scheffler.
\newblock \emph{Phys. Rev. Lett.}, 106\penalty0 (15):\penalty0 153003--4, 2011.

\bibitem[Ren et~al.(2013)Ren, Rinke, Scuseria, and Scheffler]{ren:2013A}
Xinguo Ren, Patrick Rinke, Gustavo~E. Scuseria, and Matthias Scheffler.
\newblock \emph{Physical Review B}, 88:\penalty0 035120, 2013.

\bibitem[van Aggelen et~al.(2013)van Aggelen, Yang, and Yang]{yang:2013A}
Helen van Aggelen, Yang Yang, and Weitao Yang.
\newblock \emph{Physical Review A}, 88:\penalty0 030501, 2013.

\bibitem[Bates and Furche(2013)]{furche:2013A}
Jefferson~E. Bates and Filipp Furche.
\newblock \emph{J. Chem. Phys.}, 139:\penalty0 171103, 2013.

\bibitem[Zhang and Xu(2013)]{xu:2013A}
Igor~Ying Zhang and Xin Xu.
\newblock \emph{J. Phys. Chem. Lett.}, 4:\penalty0 1669--1675, 2013.

\bibitem[gmt()]{gmtkn30}
\url{http://www.thch.uni-bonn.de/tc/index.php?section=downloads&subsection=GMTKN30&lang=english}.
\newblock Accessed: 2017-05-25.

\bibitem[m15()]{m15db}
\url{https://comp.chem.umn.edu/db2015/}.
\newblock Accessed: 2017-05-25.

\bibitem[Pople et~al.(1989)Pople, Head-Gordon, Fox, Raghavachari, and
  Curtiss]{curtiss:1989A}
J.~A. Pople, M~Head-Gordon, D.~J. Fox, K.~Raghavachari, and L.~A. Curtiss.
\newblock \emph{J. Chem. Phys.}, 90:\penalty0 5622--5629, 1989.

\bibitem[Curtiss et~al.(1991)Curtiss, Raghavachari, Trucks, and
  Pople]{curtiss:1991A}
L.~A. Curtiss, K.~Raghavachari, G.~W. Trucks, and J.~A. Pople.
\newblock \emph{J. Chem. Phys.}, 94:\penalty0 7221--7230, 1991.

\bibitem[Curtiss et~al.(1998)Curtiss, Raghavachari, Redfern, Rassolov, and
  Pople]{curtiss:1998A}
L.~A. Curtiss, K.~Raghavachari, P.~C. Redfern, V.~Rassolov, and J.~A. Pople.
\newblock \emph{J. Chem. Phys.}, 109:\penalty0 7764--7776, 1998.

\bibitem[Curtiss et~al.(2007)Curtiss, Redfern, and Raghavachari]{curtiss:2007A}
LA~Curtiss, PC~Redfern, and K~Raghavachari.
\newblock \emph{J. Chem. Phys.}, 126:\penalty0 084108--12, 2007.

\bibitem[Sinano{\v g}lu(1962)]{sinanoglu:1962A}
Oktay Sinano{\v g}lu.
\newblock \emph{J. Chem. Phys.}, 36:\penalty0 706, 1962.

\bibitem[Purvis and Bartlett(1982)]{purvis_full_1982}
George~D. Purvis and Rodney~J. Bartlett.
\newblock \emph{The Journal of Chemical Physics}, 76\penalty0 (4):\penalty0
  1910--1918, 1982.

\bibitem[Raghavachari et~al.(1989)Raghavachari, Trucks, Pople, and
  Head-Gordon]{raghavachari_fifth-order_1989}
Krishnan Raghavachari, Gary~W. Trucks, John~A. Pople, and Martin Head-Gordon.
\newblock \emph{Chemical Physics Letters}, 157\penalty0 (6):\penalty0 479--483,
  1989.

\bibitem[Booth et~al.(2009)Booth, Thom, and Alavi]{booth:2009A}
George~H. Booth, Alex J.~W. Thom, and Ali Alavi.
\newblock \emph{J. Chem. Phys.}, 131:\penalty0 054106, 2009.

\bibitem[Cleland et~al.(2010)Cleland, Booth, and Alavi]{booth:2010A}
Deidre Cleland, George~H. Booth, and Ali Alavi.
\newblock \emph{J. Chem. Phys.}, 132:\penalty0 041103, 2010.

\bibitem[Cleland et~al.(2011)Cleland, Booth, and Alavi]{booth:2011A}
D.~M. Cleland, George~H. Booth, and Ali Alavi.
\newblock \emph{J. Chem. Phys.}, 134:\penalty0 024112, 2011.

\bibitem[Delley(1990)]{delley:1990A}
B.~Delley.
\newblock \emph{J. Chem. Phys.}, 92:\penalty0 508--517, 1990.

\bibitem[Blum et~al.(2009)Blum, Gehrke, Hanke, Havu, Havu, Ren, Reuter, and
  Scheffler]{scheffler:2009A}
Volker Blum, Ralf Gehrke, Felix Hanke, Paula Havu, Ville Havu, Xinguo Ren,
  Karsten Reuter, and Matthias Scheffler.
\newblock \emph{Comput. Phys. Comm.}, 180:\penalty0 2175--2196, 2009.

\bibitem[Møller and Plesset(1934)]{moller_note_1934}
Chr. Møller and M.~S. Plesset.
\newblock \emph{Phys. Rev.}, 46\penalty0 (7):\penalty0 618--622, 1934.

\bibitem[Bohm and Pines(1951)]{bohm_collective_1951}
David Bohm and David Pines.
\newblock \emph{Phys. Rev.}, 82\penalty0 (5):\penalty0 625--634, 1951.

\bibitem[Pines and Bohm(1952)]{pines_collective_1952}
David Pines and David Bohm.
\newblock \emph{Phys. Rev.}, 85\penalty0 (2):\penalty0 338--353, 1952.

\bibitem[Bohm and Pines(1953)]{bohm_collective_1953}
David Bohm and David Pines.
\newblock \emph{Phys. Rev.}, 92\penalty0 (3):\penalty0 609--625, 1953.

\bibitem[Furche(2001)]{furche:2001A}
Filipp Furche.
\newblock \emph{Phys. Rev. B}, 64:\penalty0 195120, 2001.

\bibitem[Koch et~al.(1994)Koch, Christiansen, Kobayashi, J{\o}rgensen, and
  Helgaker]{helgaker:1994A}
Henrik Koch, Ove Christiansen, Rika Kobayashi, Poul J{\o}rgensen, and Trygve
  Helgaker.
\newblock \emph{Chem. Phys. Lett.}, 228:\penalty0 233--238, 1994.

\bibitem[Dunning(1989)]{dunning:1989A}
T.~H. Dunning.
\newblock \emph{J. Chem. Phys.}, 90:\penalty0 1007--1023, 1989.

\bibitem[Helgaker et~al.(1997)Helgaker, Klopper, Koch, and Noga]{klopper:1997A}
Trygve Helgaker, Wim Klopper, Henrik Koch, and Jozef Noga.
\newblock \emph{J. Chem. Phys.}, 106\penalty0 (23):\penalty0 9639--9646, 1997.

\bibitem[Wolinski and Pulay(2003)]{pulay:2003A}
K.~Wolinski and P.~Pulay.
\newblock \emph{J. Chem. Phys.}, 118:\penalty0 9497--9503, 2003.

\bibitem[Zhang et~al.(2010{\natexlab{b}})Zhang, Luo, and Xu]{igor:2010A}
I.~Y. Zhang, Y~Luo, and X~Xu.
\newblock \emph{J. Chem. Phys.}, 133:\penalty0 104105, 2010{\natexlab{b}}.

\bibitem[Tew et~al.(2007)Tew, Klopper, and Helgaker]{klopper:2007A}
David~P. Tew, Wim Klopper, and Trygve Helgaker.
\newblock \emph{J. Comput. Chem.}, 28:\penalty0 1307--1320, 2007.

\bibitem[Kong et~al.(2012)Kong, Bischoff, and Valeev]{kong:2012A}
Liguo Kong, Florian~A. Bischoff, and Edward~F. Valeev.
\newblock \emph{Chem. Rev.}, 112:\penalty0 75--107, 2012.

\bibitem[Kutzelnigg and Klopper(1991)]{klopper:1991A}
Werner Kutzelnigg and Wim Klopper.
\newblock \emph{J. Chem. Phys.}, 94:\penalty0 1985--2001, 1991.

\bibitem[Noga et~al.(1992)Noga, Kutzelnigg, and Klopper]{klopper:1992A}
Josef Noga, Werner Kutzelnigg, and Wim Klopper.
\newblock \emph{Chem. Phys. Lett.}, 199:\penalty0 497--504, 1992.

\bibitem[H\"attig et~al.(2012)H\"attig, Klopper, K\"ohn, and
  Tew]{klopper:2012A}
Christof H\"attig, Wim Klopper, Andreas K\"ohn, and David~P. Tew.
\newblock \emph{Chem. Rev.}, 112:\penalty0 4--74, 2012.

\bibitem[Ten-no and Noga(2012)]{noga:2012A}
Seiichiro Ten-no and Jozef Noga.
\newblock \emph{Wiley Interdiscip. Rev.: Comput. Mol. Sci.}, 2:\penalty0
  114--125, 2012.

\bibitem[Tao et~al.(2003)Tao, Perdew, Staroverov, and
  Scuseria]{tao_climbing_2003}
Jianmin Tao, John~P. Perdew, Viktor~N. Staroverov, and Gustavo~E. Scuseria.
\newblock \emph{Phys. Rev. Lett.}, 91\penalty0 (14):\penalty0 146401, 2003.

\bibitem[Staroverov et~al.(2004)Staroverov, Scuseria, Tao, and
  Perdew]{staroverov_tests_2004}
Viktor~N. Staroverov, Gustavo~E. Scuseria, Jianmin Tao, and John~P. Perdew.
\newblock \emph{Phys. Rev. B}, 69\penalty0 (7):\penalty0 075102, 2004.

\bibitem[Heyd and Scuseria(2004)]{heyd_efficient_2004}
Jochen Heyd and Gustavo~E. Scuseria.
\newblock \emph{The Journal of Chemical Physics}, 121\penalty0 (3):\penalty0
  1187--1192, 2004.

\bibitem[Paier et~al.(2006)Paier, Marsman, Hummer, Kresse, Gerber, and
  \'A{}ngy\'an]{kresse:2006A}
J.~Paier, M.~Marsman, K.~Hummer, G.~Kresse, I.~C. Gerber, and J.~G.
  \'A{}ngy\'an.
\newblock \emph{J. Chem. Phys.}, 124:\penalty0 154709, 2006.

\bibitem[Schimka et~al.(2011)Schimka, Harl, and Kresse]{kresse:2011A}
Laurids Schimka, Judith Harl, and Georg Kresse.
\newblock \emph{J. Chem. Phys.}, 134:\penalty0 024116, 2011.

\bibitem[Paier et~al.(2007)Paier, Marsman, and Kresse]{kresse:2007A}
Joachim Paier, Martijn Marsman, and Georg Kresse.
\newblock 127:\penalty0 024103, 2007.

\bibitem[Harl and Kresse(2009)]{kresse:2009A}
Judith Harl and Georg Kresse.
\newblock \emph{Phys. Rev. Lett.}, 103:\penalty0 056401--4, 2009.

\bibitem[Gr\"uneis et~al.(2010)Gr\"uneis, Marsman, and Kresse]{gruneis:2010A}
Andreas Gr\"uneis, Martijn Marsman, and Georg Kresse.
\newblock \emph{J. Chem. Phys.}, 133:\penalty0 074107, 2010.

\bibitem[Harl et~al.(2010)Harl, Schimka, and Kresse]{kresse:2010A}
Judith Harl, Laurids Schimka, and Georg Kresse.
\newblock \emph{Phys. Rev. B}, 81:\penalty0 115126--18, 2010.

\bibitem[Haas et~al.(2009{\natexlab{a}})Haas, Tran, and
  Blaha]{haas_calculation_2009}
Philipp Haas, Fabien Tran, and Peter Blaha.
\newblock \emph{Phys. Rev. B}, 79\penalty0 (8):\penalty0 085104,
  2009{\natexlab{a}}.

\bibitem[Haas et~al.(2009{\natexlab{b}})Haas, Tran, and
  Blaha]{haas_erratum:_2009}
Philipp Haas, Fabien Tran, and Peter Blaha.
\newblock \emph{Phys. Rev. B}, 79\penalty0 (20):\penalty0 209902,
  2009{\natexlab{b}}.

\bibitem[Tran et~al.(2007)Tran, Laskowski, Blaha, and
  Schwarz]{tran_performance_2007}
Fabien Tran, Robert Laskowski, Peter Blaha, and Karlheinz Schwarz.
\newblock \emph{Phys. Rev. B}, 75\penalty0 (11):\penalty0 115131, 2007.

\bibitem[Klime{\v s} et~al.(2011)Klime{\v s}, Bowler, and
  Michaelides]{klimes_van_2011}
Ji{\v r}{\' i} Klime{\v s}, David~R. Bowler, and Angelos Michaelides.
\newblock \emph{Phys. Rev. B}, 83\penalty0 (19):\penalty0 195131, 2011.

\bibitem[Csonka et~al.(2009)Csonka, Perdew, Ruzsinszky, Philipsen, Leb{\`e}gue,
  Paier, Vydrov, and {\' A}ngy{\' a}n]{csonka_assessing_2009}
G{\' a}bor~I. Csonka, John~P. Perdew, Adrienn Ruzsinszky, Pier H.~T. Philipsen,
  S{\' e}bastien Leb{\`e}gue, Joachim Paier, Oleg~A. Vydrov, and J{\' a}nos~G.
  {\' A}ngy{\' a}n.
\newblock 79\penalty0 (15):\penalty0 155107, 2009.

\bibitem[Tran et~al.(2016)Tran, Stelzl, and Blaha]{tran_rungs_2016}
Fabien Tran, Julia Stelzl, and Peter Blaha.
\newblock \emph{The Journal of Chemical Physics}, 144:\penalty0 204120, 2016.

\bibitem[Shepherd and Gr\"uneis(2013)]{gruneis:2013A}
James~J. Shepherd and Andreas Gr\"uneis.
\newblock \emph{Phys. Rev. Lett.}, 110:\penalty0 226401, 2013.

\bibitem[Booth et~al.(2013)Booth, Gr\"uneis, Kresse, and Alavi]{booth:2013A}
George~H. Booth, Andreas Gr\"uneis, Georg Kresse, and Ali Alavi.
\newblock \emph{Nature}, 493:\penalty0 365--370, 2013.

\bibitem[Shepherd et~al.(2014)Shepherd, Henderson, and
  Scuseria]{shepherd:2014A}
James~J. Shepherd, Thomas~M. Henderson, and Gustavo~E. Scuseria.
\newblock \emph{J. Chem. Phys.}, 140:\penalty0 124102, 2014.

\bibitem[Gr\"uneis(2015{\natexlab{a}})]{gruneis:2015A}
Andreas Gr\"uneis.
\newblock \emph{Phys. Rev. Lett.}, 115:\penalty0 066402, 2015{\natexlab{a}}.

\bibitem[Gr\"uneis(2015{\natexlab{b}})]{gruneis:2015B}
Andreas Gr\"uneis.
\newblock \emph{J. Chem. Phys.}, 143:\penalty0 102817, 2015{\natexlab{b}}.

\bibitem[McClain et~al.(2017)McClain, Sun, Chan, and Berkelbach]{timothy:2017A}
James McClain, Qiming Sun, Garnet Kin-Lic Chan, and Timothy~C. Berkelbach.
\newblock \emph{J. Chem. Theory Comput.}, 13:\penalty0 1209--1218, 2017.

\bibitem[Shepherd et~al.(2012)Shepherd, Booth, Gr\"uneis, and
  Alavi]{booth:2012A}
James~J. Shepherd, George Booth, Andreas Gr\"uneis, and Ali Alavi.
\newblock \emph{Phys. Rev. B}, 85:\penalty0 081103, 2012.

\bibitem[Bressanini and Reynolds(1999)]{bressanini:1999A}
D.~Bressanini and P.~J. Reynolds.
\newblock \emph{Adv. Chem. Phys.}, 105:\penalty0 37, 1999.

\bibitem[Needs et~al.(2002)Needs, Kent, Porter, Towler, and
  Rajagopal]{needs:2002A}
R.~J. Needs, P.~R.~C. Kent, A.~R. Porter, M.~D. Towler, and G.~Rajagopal.
\newblock \emph{Int. J. Quantum Chem.}, 86:\penalty0 218--225, 2002.

\bibitem[Grossman(2002)]{grossman:2002A}
J.~C. Grossman.
\newblock \emph{J. Chem. Phys.}, 117:\penalty0 1434, 2002.

\bibitem[Lester~Jr. et~al.(2009-08-17)Lester~Jr., Mitas, and
  Hammond]{mitas:2009A}
William~A. Lester~Jr., Lubos Mitas, and Brian Hammond.
\newblock \emph{Chem. Phys. Lett.}, 478:\penalty0 1--10, 2009-08-17.

\bibitem[Dubeck{\` y} et~al.(2016)Dubeck{\` y}, Mitas, and Jure{\v
  c}ka]{mitas:2016A}
Mat{\' u}{\v s} Dubeck{\` y}, Lubos Mitas, and Petr Jure{\v c}ka.
\newblock \emph{Chem. Rev.}, 116:\penalty0 5188--5215, 2016.

\bibitem[Chadi and Cohen(1973)]{chadi:1973A}
D.J. Chadi and M.~L. Cohen.
\newblock \emph{Phys. Rev. B}, 8:\penalty0 5747, 1973.

\bibitem[Marcus et~al.(1971)Marcus, Janak, and Williams]{marcus:1971}
P.~P. Marcus, J.~F. Janak, and A.~R. Williams.
\newblock volume~4 of \emph{10}.
\newblock Plenum Press, New York, 1971.
\newblock ISBN 978-1-4684-1892-7.

\bibitem[Baldereschi(1973)]{baldereschi:1973A}
A.~Baldereschi.
\newblock \emph{Phys. Rev. B}, 7:\penalty0 5212, 1973.

\bibitem[Methfessel and Paxton(1989)]{methfessel:1989A}
M.~Methfessel and A.~T. Paxton.
\newblock \emph{Phys. Rev. B}, 40:\penalty0 4616, 1989.

\bibitem[Bl{\"o}chl(1994)]{bloechl:1994A}
P.~E. Bl{\"o}chl.
\newblock \emph{Phys. Rev. B}, 49:\penalty0 16223, 1994.

\bibitem[Gr\"uneich and He\ss(1998)]{gruneich:1998A}
Armin Gr\"uneich and Bernd~A. He\ss.
\newblock \emph{Theor. Chem. Accounts}, 100:\penalty0 253--263, 1998.

\bibitem[Erba and Halo(2009)]{cryscor09:2009A}
A.~Erba and M.~Halo.
\newblock Cryscor09 \emph{User's Manual}, university of torino, 2009.

\bibitem[Hellmann()]{hellmann:1935A}
H.~Hellmann.
\newblock \emph{J. Chem. Phys.}, 3:\penalty0 61.

\bibitem[Schwerdtfeger(2011)]{schwerdtfeger:2011A}
Peter Schwerdtfeger.
\newblock \emph{ChemPhysChem}, 12:\penalty0 3143--3155, 2011.

\bibitem[Levchenko et~al.(2015)Levchenko, Ren, Wieferink, Johanni, Rinke, Blum,
  and Scheffler]{levchenko:2015A}
Sergey~V. Levchenko, Xinguo Ren, J{\" u}rgen Wieferink, Rainer Johanni, Patrick
  Rinke, Volker Blum, and Matthias Scheffler.
\newblock \emph{Comput. Phys. Comm.}, 192:\penalty0 60--69, 2015.

\bibitem[Singh and Nordstr\"om(2006)]{LAPW:2006A}
David~J. Singh and Lars Nordstr\"om.
\newblock \emph{Planewaves, Pseudopotentials, and the LAPW Method}.
\newblock Splinger, New York, NY, 2006.

\bibitem[Lejaeghere et~al.(2014)Lejaeghere, Speybroeck, Oost, and
  Cottenier]{cottenier:2014A}
K.~Lejaeghere, V.~Van Speybroeck, G.~Van Oost, and S.~Cottenier.
\newblock \emph{Critical Reviews in Solid State and Materials Sciences},
  39\penalty0 (1), 2014.

\bibitem[wie()]{wien2k:2014A}
\url{http://molmod.ugent.be/DeltaCodesDFT}.
\newblock Accessed: 2015-08-17.

\bibitem[Blaha et~al.(1999)Blaha, Schwzrz, Madsen, Kvasnicka, and
  Luitz]{wien2k:1999A}
P.~Blaha, K.~Schwzrz, G.~K.~H. Madsen, D.~Kvasnicka, and J.~Luitz.
\newblock Wien2k, anaugmented plane wave + local orbitals program for
  calculating crystal properties, karlheinz schwart, techn. universit\"at wien,
  austria, 1999.

\bibitem[Zhao and Truhlar(2008)]{truhlar:2008A}
Y.~Zhao and D.~G. Truhlar.
\newblock \emph{Theor. Chem. Acc.}, 120:\penalty0 215--241, 2008.

\bibitem[Sun et~al.(2015)Sun, Ruzsinszky, and Perdew]{perdew:2015A}
Jianwei Sun, Adrienn Ruzsinszky, and John P. Perdew.
\newblock \emph{Phys. Rev. Lett.}, 115:\penalty0 036402, 2015.

\bibitem[Krukau et~al.(2006)Krukau, Vydrov, Izmaylov, and
  Scuseria]{krukau_influence_2006}
Aliaksandr~V. Krukau, Oleg~A. Vydrov, Artur~F. Izmaylov, and Gustavo~E.
  Scuseria.
\newblock \emph{J. Chem. Phys.}, 125\penalty0 (22):\penalty0 224106, 2006.

\bibitem[Ren et~al.(2012{\natexlab{b}})Ren, Rinke, Joas, and
  Scheffler]{ren:2012B}
Xinguo Ren, Patrick Rinke, Christian Joas, and Matthias Scheffler.
\newblock \emph{J. Mater. Sci.}, 47:\penalty0 7447--7471, 2012{\natexlab{b}}.

\bibitem[Pisani et~al.(2008)Pisani, Maschio, Casassa, Halo, Sch\"utz, and
  Usvyat]{pisani:2008A}
Cesare Pisani, Lorenzo Maschio, Silvia Casassa, Migen Halo, Martin Sch\"utz,
  and Denis Usvyat.
\newblock \emph{J. Comput. Chem.}, 29:\penalty0 2113--2124, 2008.

\bibitem[Pisani et~al.(2012)Pisani, Sch\"utz, Casassa, Usvyat, Maschio, Lorenz,
  and Erba]{pisani:2012A}
Cesare Pisani, Martin Sch\"utz, Silvia Casassa, Denis Usvyat, Lorenzo Maschio,
  Marco Lorenz, and Alessandro Erba.
\newblock \emph{Phys. Chem. Chem. Phys.}, 14:\penalty0 7615--7628, 2012.

\bibitem[Del~Ben et~al.(2012)Del~Ben, Hutter, and VandeVondele]{mauro:2012A}
Mauro Del~Ben, J\"urg Hutter, and Joost VandeVondele.
\newblock \emph{J. Chem. Theory Comput.}, 8:\penalty0 4177--4188, 2012.

\bibitem[Del~Ben et~al.(2013)Del~Ben, Hutter, and VandeVondele]{mauro:2013A}
M.~Del~Ben, J.~Hutter, and J.~VandeVondele.
\newblock \emph{J. Chem. Theory Comput.}, 9:\penalty0 2654, 2013.

\bibitem[Zhang et~al.(2016)Zhang, Rinke, Perdew, and Scheffler]{igor:2016A}
Igor~Ying Zhang, Patrick Rinke, John~P. Perdew, and Matthias Scheffler.
\newblock \emph{Phys. Rev. Lett.}, 117:\penalty0 133002, 2016.

\bibitem[Whitten(1973)]{whitten:1973A}
J.~L. Whitten.
\newblock \emph{J. Chem. Phys.}, 58:\penalty0 4496, 1973.

\bibitem[Dunlap(2000)]{dunlap:2000A}
B.~Dunlap.
\newblock \emph{J. Mol. Struct.: THEOCHEM}, 501:\penalty0 221, 2000.

\bibitem[Ihrig et~al.(2015)Ihrig, Wieferink, Zhang, Ropo, Ren, Rinke,
  Scheffler, and Blum]{igor:2015A}
Arvid~Conrad Ihrig, J{\" u}rgen Wieferink, Igor~Ying Zhang, Matti Ropo, Xinguo
  Ren, Patrick Rinke, Matthias Scheffler, and Volker Blum.
\newblock \emph{New J. Phys.}, 17:\penalty0 093020, 2015.

\bibitem[Ayala et~al.(2001)Ayala, Kudin, and Scuseria]{scuseria:2001A}
PY~Ayala, KN~Kudin, and GE~Scuseria.
\newblock \emph{J. Chem. Phys.}, 115:\penalty0 9698--9707, 2001.

\bibitem[Doser et~al.(2010)Doser, Zienau, Clin, Lambrecht, and
  Ochsenfeld]{oschsenfeld:2010A}
Bernd Doser, Jan Zienau, Lucien Clin, Daniel~S. Lambrecht, and Christian
  Ochsenfeld.
\newblock \emph{Z. Phys. Chem.}, 224:\penalty0 397--412, 2010.

\bibitem[Lejaeghere et~al.(2016)Lejaeghere, Bihlmayer, Bj{\" o}rkman, Blaha,
  Bl{\" u}gel, Blum, Caliste, Castelli, Clark, Corso, Gironcoli, Deutsch,
  Dewhurst, Marco, Draxl, Dulak, Eriksson, Flores-Livas, Garrity, Genovese,
  Giannozzi, Giantomassi, Goedecker, Gonze, Gr{\aa}n{\" a}s, Gross, Gulans,
  Gygi, Hamann, Hasnip, Holzwarth, Iu{\c s}an, Jochym, Jollet, Jones, Kresse,
  Koepernik, K{\" u}c{\" u}kbenli, Kvashnin, Locht, Lubeck, Marsman, Marzari,
  Nitzsche, Nordstr{\" o}m, Ozaki, Paulatto, Pickard, Poelmans, Probert,
  Refson, Richter, Rignanese, Saha, Scheffler, Schlipf, Schwarz, Sharma,
  Tavazza, Thunstr{\" o}m, Tkatchenko, Torrent, Vanderbilt, Setten, Speybroeck,
  Wills, Yates, Zhang, and Cottenier]{cottenier:2016B}
Kurt Lejaeghere, Gustav Bihlmayer, Torbj{\" o}rn Bj{\" o}rkman, Peter Blaha,
  Stefan Bl{\" u}gel, Volker Blum, Damien Caliste, Ivano~E. Castelli,
  Stewart~J. Clark, Andrea~Dal Corso, Stefano~de Gironcoli, Thierry Deutsch,
  John~Kay Dewhurst, Igor~Di Marco, Claudia Draxl, Marcin Dulak, Olle Eriksson,
  Jos{\' e}~A. Flores-Livas, Kevin~F. Garrity, Luigi Genovese, Paolo Giannozzi,
  Matteo Giantomassi, Stefan Goedecker, Xavier Gonze, Oscar Gr{\aa}n{\" a}s,
  E.~K.~U. Gross, Andris Gulans, Francois Gygi, D.~R. Hamann, Phil~J. Hasnip,
  N.~a.~W. Holzwarth, Diana Iu{\c s}an, Dominik~B. Jochym, Francois Jollet,
  Daniel Jones, Georg Kresse, Klaus Koepernik, Emine K{\" u}c{\" u}kbenli,
  Yaroslav~O. Kvashnin, Inka L.~M. Locht, Sven Lubeck, Martijn Marsman, Nicola
  Marzari, Ulrike Nitzsche, Lars Nordstr{\" o}m, Taisuke Ozaki, Lorenzo
  Paulatto, Chris~J. Pickard, Ward Poelmans, Matt I.~J. Probert, Keith Refson,
  Manuel Richter, Gian-Marco Rignanese, Santanu Saha, Matthias Scheffler,
  Martin Schlipf, Karlheinz Schwarz, Sangeeta Sharma, Francesca Tavazza, Patrik
  Thunstr{\" o}m, Alexandre Tkatchenko, Marc Torrent, David Vanderbilt, Michiel
  J.~van Setten, Veronique~Van Speybroeck, John~M. Wills, Jonathan~R. Yates,
  Guo-Xu Zhang, and Stefaan Cottenier.
\newblock \emph{Science}, 351:\penalty0 aad3000, 2016.

\bibitem[De~Waele et~al.(2016)De~Waele, Lejaeghere, Sluydts, and
  Cottenier]{cottenier:2016A}
Sam De~Waele, Kurt Lejaeghere, Michael Sluydts, and Stefaan Cottenier.
\newblock \emph{Phys. Rev. B}, 94:\penalty0 235418, 2016.

\bibitem[Pascal~Pernot and Savin(2015)]{savin:2015A}
Davide~Presti Pascal~Pernot, Bartolomeo~Civalleri and Andreas Savin.
\newblock \emph{J. Phys. Chem. A}, 119:\penalty0 5288--5304, 2015.

\bibitem[Adrienn and John~P.(2010)]{adrienn:2010A}
Ruzsinszky Adrienn and Perdew John~P.
\newblock \emph{J. Chem. Theory Comput.}, 6:\penalty0 127--134, 2010.

\bibitem[He\ss{}elmann and G\"orling(2010)]{gorling:2010A}
Andreas He\ss{}elmann and Andreas G\"orling.
\newblock \emph{Mol. Phys.}, 108:\penalty0 359--372, 2010.

\bibitem[Sun et~al.(2013)Sun, Xiao, Fang, Haunschild, Hao, Ruzsinszky, Csonka,
  Scuseria, and Perdew]{perdew:2013A}
Jianwei Sun, Bing Xiao, Yuan Fang, Robin Haunschild, Pan Hao, Adrienn
  Ruzsinszky, G{\' a}bor~I. Csonka, Gustavo~E. Scuseria, and John~P. Perdew.
\newblock \emph{Phys. Rev. Lett.}, 111:\penalty0 106401, 2013.

\bibitem[Monkhorst and Pack(1976)]{monkhorst:1976A}
Hendrik~J. Monkhorst and James~D. Pack.
\newblock \emph{Phys. Rev. B}, 13:\penalty0 5188--5192, 1976.

\bibitem[aims team(2015)]{FHI-aims:2015A}
FHI\ aims team.
\newblock Fhi-aims \emph{User's Guide}, fritz-haber-institut der
  max-planck-gesellschaft, 2015.

\bibitem[VAS()]{VASP:2015A}
\url{http://cms.mpi.univie.ac.at/vasp/vasp/Automatic_k_mesh_generation.html}.
\newblock Accessed: 2015-08-17.

\bibitem[Gr\"uneis et~al.(2011)Gr\"uneis, Booth, Marsman, Spencer, Alavi, and
  Kresse]{gruneis:2011A}
Andreas Gr\"uneis, George~H. Booth, Martijn Marsman, James Spencer, Ali Alavi,
  and Georg Kresse.
\newblock \emph{J. Chem. Theory Comput.}, 7:\penalty0 2780--2785, 2011.

\bibitem[Spencer and Alavi(2008)]{alavi:2008A}
James Spencer and Ali Alavi.
\newblock \emph{Phys. Rev. B}, 77:\penalty0 193110, 2008.

\bibitem[Marsman et~al.(2009)Marsman, Gr{\"u}neis, Paier, and
  Kresse]{gruneis:2009A}
M.~Marsman, A.~Gr{\"u}neis, J.~Paier, and G.~Kresse.
\newblock \emph{J. Chem. Phys.}, 130:\penalty0 184103, 2009.

\bibitem[Spackman and Mitchell(2001)]{spackman:2001A}
Mark~A. Spackman and Anthony~S. Mitchell.
\newblock \emph{Phys. Chem. Chem. Phys.}, 3:\penalty0 1518--1523, 2001.

\bibitem[Champagne et~al.(2003)Champagne, Jacquemin, Gu, Aoki, Kirtman, and
  Bishop]{champagne:2003A}
Benoı̂t Champagne, Denis Jacquemin, Feng~Long Gu, Yuriko Aoki, Bernard
  Kirtman, and David~M. Bishop.
\newblock \emph{Chem. Phys. Lett.}, 373:\penalty0 539--549, 2003.

\bibitem[Dovesi et~al.(2013)Dovesi, Saunders, Roetti, Orlando, Zicovich-Wilson,
  Pascale, Civalleri, Doll, Harrion, Bush, D'Arco, and
  Llundll]{crystal09:2013A}
R.~Dovesi, V.R. Saunders, C.~Roetti, R.~Orlando, C.~M. Zicovich-Wilson,
  F.~Pascale, B.~Civalleri, K.~Doll, N.M. Harrion, I.J. Bush, Ph. D'Arco, and
  M.~Llundll.
\newblock Crystal09 v2.0.1 \emph{User's Manual}, 2013.

\bibitem[Johnson et~al.(2009)Johnson, Becke, Sherrill, and
  DiLabio]{becke:2009A}
Erin~R. Johnson, Axel~D. Becke, C.~David Sherrill, and Gino~A. DiLabio.
\newblock \emph{J. Chem. Phys.}, 131:\penalty0 034111--7, 2009.

\bibitem[Gräfenstein et~al.(2007)Gräfenstein, Izotov, and
  Cremer]{cremer:2007A}
Jürgen Gräfenstein, Dmitry Izotov, and Dieter Cremer.
\newblock \emph{J. Chem. Phys.}, 127:\penalty0 214103, 2007.

\bibitem[Ohnishi and Hirata(2010)]{hirata:2010A}
Yu-ya Ohnishi and So~Hirata.
\newblock \emph{J. Chem. Phys.}, 133:\penalty0 034106, 2010.

\bibitem[Wiberg(2004)]{wiberg:2004A}
Kenneth~B. Wiberg.
\newblock \emph{J. Comput. Chem.}, 25:\penalty0 1342--1346, 2004.

\bibitem[Kutzelnigg and Morgan(1992)]{morgan:1992A}
Werner Kutzelnigg and John~D. Morgan.
\newblock \emph{J. Chem. Phys.}, 96:\penalty0 4484--4508, 1992.

\bibitem[Feller and Sordo(2000)]{feller:2000A}
David Feller and Jose~A. Sordo.
\newblock \emph{J. Chem. Phys.}, 113:\penalty0 485--493, 2000.

\end{thebibliography}
\end{document}